\pdfoutput=1
\documentclass[a4paper,12pt]{article}

\usepackage{amsmath,amsthm,amssymb,bm} 
\usepackage{graphicx}

\allowdisplaybreaks[3]

\newcommand{\ti}[1]{\tilde{#1}}     
\newcommand{\f}[2]{{\frac{#1}{#2}}} 
\newcommand{\s}[1]{\sqrt{#1}}       
\newcommand{\sfrac}[2]{{#1/#2}}     
\def\l{\left}  
\def\r{\right} 
\newcommand{\fgref}[1]{{\figurename\;\ref{#1}}} 
\newcommand{\abs}[1]{{\left|#1\right|}}  
\newcommand{\bR}{\mathbb{R}} 
\newcommand{\cN}{\mathcal{N}} 
\newcommand{\cR}{\mathcal{R}} 
\newcommand{\cW}{\mathcal{W}} 
\newcommand{\cC}{\mathcal{C}} 
\newcommand{\SO}{\mathrm{SO}} 
\newcommand{\SU}{\mathrm{SU}} 
\newcommand{\U}{\mathrm{U}}   
\newcommand{\hyphen}{{\mathchar"712D}} 
\begin{document}

\begin{titlepage}
\thispagestyle{empty}
\begin{flushright}
OU-HET-843,
RIKEN-MP-100,
RIKEN-QHP-175
\end{flushright}

\bigskip

\begin{center}
{\Large 
Holographic Heavy Quark Symmetry
}
\end{center}

\bigskip

\begin{center}

Koji {\sc Hashimoto}$^{\spadesuit\heartsuit,}$%
\footnote{E-mail :  koji@phys.sci.osaka-u.ac.jp},
Noriaki {\sc Ogawa}$^{\heartsuit\diamondsuit,}$%
\footnote{E-mail : noriaki@riken.jp}
and
Yasuhiro {\sc Yamaguchi}$^{\heartsuit,}$%
\footnote{E-mail : yasuhiro.yamaguchi@riken.jp}

\setcounter{footnote}{0}

\bigskip

${}^\spadesuit$
{\it
Department of Physics, Osaka University, \\
Toyonaka, Osaka 560-0043, Japan
}\\

\bigskip

${}^\heartsuit$
{\it
Mathematical Physics Laboratory,
RIKEN Nishina Center, \\
Wako, Saitama 351-0198, Japan
}\\

\bigskip

${}^\diamondsuit$
{\it
Quantum Hadron Physics Laboratory,
RIKEN Nishina Center, \\
Wako, Saitama 351-0198, Japan
}

\end{center}

\bigskip

\begin{abstract}
We investigate the heavy quark spin symmetry,
i.e. the mass degeneracy of pseudo-scalar
and vector quarkonia at heavy quark limit,
by using the gauge/gravity correspondence.
We allow generic D3-like geometry
with a flavor D7-brane, to avoid supersymmetric mass degeneracy.
For geometries admitting physical QCD-like properties,
we find that the mass degeneracy is generically
achieved in a good accuracy, up to a few percent mass splitting.
We also compute spectra of excited quarkonia states,
and discuss comparisons with experiments
and quark-model calculations.
\end{abstract}

\end{titlepage}

\tableofcontents

\section{Introduction}

Hadrons containing heavy quarks attract a great deal of interest in hadron and
nuclear physics.
Out of the six flavors of quarks, the charm, bottom and top quarks are
classified as a heavy quark because they
have a mass which is much larger than the typical scale of
the Quantum Chromodynamics (QCD), $\Lambda_{\rm QCD}$. 
There are active researches of hadrons with charm and/or bottom
quarks,
while hadrons containing top quarks are not found because 
a top quark is easy to decay via the weak
interaction.
Accelerator experiments have found 
conventional hadrons which are described by the simple quark model
expressing baryons as $qqq$ states 
and mesons as $q\bar{q}$ states with constituent (anti-)quarks
$q(\bar{q})$~\cite{Agashe:2014kda,Eichten:2007qx,Klempt:2009pi}.
In addition,
interesting observations of 
exotic quarkonia, called $X,Y,Z$ states, also motivate us to study the
hadron spectrum in the heavy flavor sector.
Discovered exotic hadrons
are expected to have complex structures such as multiquark states and
hadron composite states~\cite{swanson:2006st,Voloshin:2007dx,Godfrey:2008nc,Brambilla:2010cs}.
The various structures of heavy hadrons are generated by the natures of internal quark
potentials and/or hadron-hadron interactions which result from
fundamental phenomena of QCD.

In the rich spectra of heavy hadrons,
a new symmetry which does not emerge in the light quark sector is considered
to be important.
This is called heavy quark
symmetry~\cite{Isgur:1989vq,Isgur:1989ed,Isgur:1991wq,Neubert:1993mb,Manohar:2000dt}.
It possesses flavor and spin symmetries for heavy quarks.
In particular, the spin symmetry leads to a specific feature which is
mass degeneracy of heavy hadrons having a different spin~\cite{Neubert:1993mb,Manohar:2000dt}.
In the case of mesons, experimental results show small mass splittings between
pseudo-scalar mesons with spin-0 and vector mesons with spin-1,
e.g., the mass splitting of $BB^\ast$ ($\sim$45 MeV) in the bottom quark sector is
much smaller than that of $KK^\ast$ ($\sim400$ MeV) in the strange quark
sector.
The mass degeneracy is also realized in the heavy baryons~\cite{Roberts:2007ni} and is
expected in the multi-hadron states such as hadronic molecules and heavy
mesonic nuclei~\cite{Yasui:2013vca,Yamaguchi:2014era}.
Furthermore, the heavy quark spin symmetry affects the heavy hadron
decays and productions~\cite{Isgur:1991wq,Manohar:2000dt,Ohkoda:2012rj,Guo:2014qra,Yasui:2014cwa}.
This symmetry provides the relations among various decay and production
ratios of heavy hadrons with different spins.
Hence, in spectroscopy of heavy hadrons,
the heavy quark spin symmetry plays a significant role.

On this symmetry, some thoretical approaches have been developed.
It is explained by suppression of the spin-dependent forces between
heavy quarks.
For hadrons including a single heavy quark,
there have been discussions about the heavy quark spin symmetry in the heavy quark
effective theory
(HQET)~\cite{Neubert:1993mb,Manohar:2000dt,Grinstein:1990mj,Georgi:1990um,Mannel:1991mc}.
HQET is given in $1/m_Q$-expansion, where $m_Q$ is the mass of the
heavy quark $Q$.
In the HQET Lagrangian, 
the spin-dependent operators are included in the higher order terms,
while the leading term is spin-independent.
Hence the heavy quark spin symmetry emerges in
the heavy quark limit $m_Q\rightarrow \infty$.
The suppression of the spin-dependent force is also expected in the
phenomenological constituent quark model.
The masses of hadrons with different spins are split by the hyperfine
interaction depending on quark spins~\cite{Godfrey:1985xj}.
This interaction between two quarks is suppressed in the heavy quark limit
because it is inversely proportional to the product of the quark masses.
The $Q\bar{Q}$ potential is also investigated by the effective
field theory
that is the potential nonrelativistic QCD (pNRQCD)~\cite{Brambilla:2004jw}.
The spin-dependent force included in the relativistic corrections has been
studied in pNRQCD by making use of lattice QCD simulations within the quench
approximation~\cite{Brambilla:2010cs,Koma:2006fw,Koma:2012bc}.

The spectra of heavy hadrons have also been studied, in various
theoretical ways~\cite{Eichten:2007qx,Klempt:2009pi}.
The constituent quark model has been applied to the hadron spectra 
from light to heavy quark sectors~\cite{Godfrey:1985xj}.
The spectroscopy of heavy quarkonia 
in perturbative QCD has made progress, too,
where the properties of heavy quarkonia 
are computed in systematic
expansions with respect to the strong coupling constant $\alpha_s$
\cite{Brambilla:2010cs,Kiyo:2013aea,Kiyo:2014uca}.
To go beyond the perturbation, 
lattice QCD is
an efficient tool for calculations of the hadron properties.
Recently several collaborations challenge the lattice QCD simulations of
hadron spectra at or closed to the physical point even in the charm and
bottom quark regions~\cite{Aoki:2013ldr}.

In addition to them, 
we find that the gauge/gravity correspondence \cite{Maldacena:1997re} is yet another promising approach,
since it provides powerful methods to deal with strongly coupled
theories. 
It has been applied to
investigate hadron spectra, by
introducing dynamical quarks 
which are described by excitations on probe D-branes.
For example, 
$\cN=2$ hypermultiplet flavors (= quark multiplets) added to $\cN=4$ $\SU(N_c)$ super Yang-Mills theory are realized
by introducing $N_f$ probe D7-branes 
on the $\mathrm{AdS}_5\times\mathrm{S}^5$ background (which is generated by $N_c$ D3-branes)
on the gravity side~\cite{Karch:2002sh}.
In this model, from the stringy point of view, the quarks are expressed as fundamental strings stretched
between D3- and D7-branes.
The configuration is utilized in the calculations of the meson spectrum.
The masses of pseudo-scalar and vector mesons are computed as the
fluctuations of the scalar and vector fields on the flavor D-branes~\cite{Kruczenski:2003be}.
The study of the meson spectrum has been investigated not only on this D3/D7
model, but also on the D4/D6 model, D4/D8 model, etc.

In the gauge theory holding the supersymmetry, however,
both pseudo-scalar and vector mesons are members of the same multiplet,
and the masses are completely degenerate regardless of the
value of the quark mass.
Even when the supersymmetry is broken by finite temperature or Shark-Schwartz compactification, etc,
the supersymmetry recovers in the heavy quark limit (i.e., in the UV limit) in most of the top-down models.
Hence, the presence of the heavy quark spin symmetry has not been obvious
in the gauge/gravity duality.
In order to see 
the heavy quark spin symmetry exists or not,
we need to investigate meson mass degeneracies
on theories
which are non-supersymmetric 
even in the UV region.

In this paper,
we propose a semi-bottom-up, deformed D3-D7 model.
The background geometry is deformed from the conventional 
$\mathrm{AdS}_5\times\mathrm{S}^5$,
and holds no supersymmetry or conformal symmetry generally.
We introduce $4$ deformation parameters for the background,
but they are constrained by several conditions so that
the theory would have physically reasonable properties.
We investigate the spectra of the pseudo-scalar and vector quarkonia
at $m_Q\to\infty$ limit,
computed as the fluctuations of the fields on the flavor D7-brane
put on this background. 
Finally, we will find that the heavy quark spin symmetry is
{\it approximatively} true, but {\it not exact} in general.
We observe a slight difference, at most $\simeq 1.5\%$.

The remaining part of
this paper is organized as follows.
Section~\ref{ch:Setup} provides the setup of our model,
including the background geometry and
the DBI action in string theory.
In Section~\ref{ch:Constraints_para},
we consider several physical conditions on that gravitational background,
and successfully determine the limited region in the parameter space
which we can focus on.
In Section~\ref{ch:EOM},
the equations of motion on the probe D7 brane are derived.
The asymptotic solutions of the equations are also given.
In Section~\ref{ch:Numerical_results},
we solve those equations of motion numerically
and obtain the mass spectra of the quarkonia.
We discuss the aspects of the results and compare with those of effective models as well as experiments.
Finally in Section~\ref{ch:Disscussion_Conclusion},
we summarize the paper and
discuss the implications of our results and possible applications.

\section{Deformed D3-D7 Model}
\label{ch:Setup}

As we explained above,
we need to study non-supersymmetric models with a flavor, 
for investigations on heavy quark symmetry.
As a simple but interesting model with such properties,
we propose a semi-bottom-up model which we call deformed D3-D7 model.
It is based on the type IIB superstring and similar to the standard D3-D7 model,
but we assume the existence of some (unknown) additional matter or flux configurations.
It leads to a backreaction to deform the D3 background geometry from $\mathrm{AdS}_5\times\mathrm{S}^5$.
We then introduce a probe D7-brane on this background, obeying the conventional DBI action.
We further assume that the couplings between the D7-brane and the unknown background fluxes 
are always suppressed.
Then we can compute the energy of the excitation modes on this D7-brane, 
which correspond to meson excitations on the dual field theory.

\subsection{Background Geometry and Probe D7-Brane}
On the gravitational background geometry,
we impose a symmetry of $\SO(1,3)\times\SO(6)$,
and adopt an ansatz for the UV ($r\gg{}^\exists r_{\mathit{IR}}$) leading form as
\begin{subequations}
\label{eq:background}
\begin{align}
\label{eq:metric:str}
  ds_{\mathit{str}}^2 &= {r}^{2\alpha}\eta_{\mu\nu}dx^{\mu}dx^{\nu}
  + R^2r^{-2\beta}\l(\f{dr^2}{r^2} + r^{2\delta}d\Omega_5^2\r)\,,\\
\label{eq:dilaton}
  e^{\phi} &= g_0\, r^{-4\gamma}\,,
\end{align}
\end{subequations}
for which the Einstein equation is maintained
by some unknown matter and flux configurations which are omitted here.
The $d\Omega_5^2$ represents the metric of $S^5$,
on which we span coordinates as
\begin{align}
\label{eq:S5coord}
  d\Omega_5^2 = d\theta^2 + \cos^2\theta d\psi^2 + \sin^2\theta d\Omega_3^2\,,
\qquad
  d\Omega_3^2 = d\sigma_1^2 + \sin^2\sigma_1 d\Omega_2^2\,.
\end{align}
This geometry has four parameters $(\alpha,\beta,\gamma,\delta)$,
and the standard D3 background corresponds to $(\alpha,\beta,\gamma,\delta)=(1,0,0,0)$.
The $10$d Einstein-frame metric is written as
\begin{align}
\label{eq:metric:Ein}
  ds_{\mathit{Ein}}^2 &= e^{-\phi/2}ds_{\mathit{str}}^2
  = g_0^{-1/2}\l[r^{2\ti\alpha}\eta_{\mu\nu}dx^{\mu}dx^{\nu}
  + R^2r^{-2\ti\beta}\l(\f{dr^2}{r^2} + r^{2\delta}d\Omega_5^2\r)\r]\,,
\end{align}
where 
\begin{align}
  \ti{\alpha}=\alpha + \gamma\,,
\qquad
  \ti{\beta}=\beta - \gamma\,.
\end{align}

We then put a probe D7-brane on this background.
It is described by the conventional DBI action
\begin{align}
\label{eq:DBI}
  S_{D7} = -T_7\int\!d^8\xi\, e^{-\phi}\s{-\det\l(h_{ab} + 2\pi\alpha'F_{ab}\r)}\,,
\end{align}
where 
$T_7$ is the D7 tension,
$h_{ab}$ is the induced (string-frame) metric on D7 and 
$F_{ab}=\partial_aA_b-\partial_bA_a$ is the field strength of the world-volume $\U(1)$ gauge field $A_{a}$.
We omitted the contributions from background fluxes, 
as was stated above.

This D7-brane corresponds to the quark sector of the theory.
It extends along with the boundary $4$-dimensional directions
and wraps on the $S^3$ in the $S^5$, which corresponds to the $\Omega_3$
in \eqref{eq:S5coord}.
That is, the configuration of the D7-branes are described by the two embedding profile functions
\begin{align}
  \theta = \Theta(r, x^\mu, \Omega_3)\,,
\qquad
  \psi = \Psi(r, x^\mu, \Omega_3)\,.
\end{align}
In particular,
we will consider static and uniform configurations
\begin{align}\label{eq:setup:D7background}
  \theta &= \Theta(r)\,,
\quad
  \psi = 0\,,
\quad
  A_{a} = 0\,,
\end{align}
and small fluctuations around it,
to describe the meson spectrum of the theory.
The D7 has a turning point at $r=r_0$ and reaches to the boundary $r\to\infty$.
In order for the continuity and smoothness at the turning point, it should satisfy
\begin{align}\label{eq:Theta:smoothness}
  \Theta(r_0)=0\,,
\qquad
  \Theta'(r_0)=\infty\,.
\end{align}

\subsection{Rescaling Symmetries}
\label{sec:rescaling}
This deformed D3-D7 model has different kinds of rescaling symmetries, 
depending on $\delta\ne 0$ or $\delta=0$.

\subsubsection*{Case 1:~$\bm{\delta\ne 0}$}
When $\delta\ne 0$,
we can rescale the exponent parameters $(\alpha,\beta,\gamma,\delta)$ 
by a positive rescaling parameter $c$ as
\begin{align}
\label{eq:rescaling:parameters}
  (\alpha,\beta,\gamma,\delta) \to c\,(\alpha,\beta,\gamma,\delta)\,,
\end{align}
by the transformation
\begin{align}
\label{eq:rescaling:coordinates}
  r\to c^{\frac{1}{\delta}}r^c\,,
\quad
  x\to c^{-\frac{\alpha}{\delta}}x\,,
\quad
  R\to c^{\frac{\beta-\delta}{\delta}}R\,.
\end{align}
We can set $\delta=0$ or $\pm 1$ by using this transformation with $c=\sfrac{1}{\abs{\delta}}$,
but we will work on general values for $\delta$ for a while.
Instead, we define
\begin{align}
  (\hat\alpha,\hat\beta,\hat{\gamma})
  = 
  \begin{cases}
  (\ti\alpha,\ti\beta,\gamma)
  &\quad(\text{for}\;\; \delta=0)\,,\\
  (\sfrac{\ti\alpha}{\abs{\delta}},\sfrac{\ti\beta}{\abs{\delta}},\sfrac{\gamma}{\abs{\delta}})
  &\quad(\text{for}\;\; \delta\ne 0)\,,
  \end{cases}
\end{align}
and sometimes use these hereafter.

\subsubsection*{Case 2:~$\bm{\delta=0}$}
When $\delta=0$, we can consider another type of scaling transformations of coordinates instead of \eqref{eq:rescaling:coordinates}, as
\begin{align}\label{eq:rescaling:delta=0}
  r\to c\, r\,, 
\quad
  x\to c^{-(\alpha+\beta)}x\,.
\end{align}
This results in a Weyl transformation
\begin{align}
  g_{MN}\to c^{-2\beta}g_{MN}\,,
\quad
  e^{\phi}\to c^{-4\gamma}e^{\phi}\,.
\end{align}
Furthermore, if we consider the redefinition of the D7 world-volume $U(1)$ gauge field
\begin{align}\label{eq:A:redef}
  A_a \to c^{-2\beta}A_a\,
\end{align}
at the same time,
we notice that the 
DBI action \eqref{eq:DBI} is kept to be invariant
up to a change of the overall coefficient.
Therefore, the set of the transformations \eqref{eq:rescaling:delta=0} and \eqref{eq:A:redef}
is a symmetry of the equations of motion for the probe D7-brane.
Since the meson spectra are determined by nothing but those EoMs,
this symmetry will take a very important roll later
(\S\ref{sec:mass_prop}, \S\ref{sec:hiddensym}).

\section{Constraints for the Parameters}
\label{ch:Constraints_para}
Although we introduced the deformed D3-D7 model in rather general form,
it proves that this model does not always corresponds to a consistent, QCD-like theory.
In this section, we find several physical conditions which we need for our purpose, 
and translate them to various constraints on the parameters $(\alpha,\beta,\gamma,\delta)$ of our model.

We will explain the essence and results for those conditions below.
The details of the derivations are given in Appendix~\ref{ch:constraints}.

\subsection{Stability and Locality}
First we consider general conditions for quantum field theories,
which do not refer to the detail of the model.
Namely, we need {\it stability} and {\it locality}.
On the gravity side,
they can be discussed for the dominating part of the theory (i.e., the gluon sector),
and are translated to so-called null energy condition
and the area-law of holographic entanglement entropy.
This is a powerful method which is applicable to holographies for various systems.
For example, non-Fermi liquid system was analyzed 
in a similar way in \cite{Ogawa2012bz}.

\subsubsection{Stability: null energy condition}
On the gravity side, the stability of the vacuum implies the
{\it null energy condition}, 
\begin{align}\label{eq:NEC:0}
  T_{MN}\xi^{M}\xi^{N} \ge 0\,,
\end{align}
imposed for the background geometry.
Here, $T_{MN}$ and $\xi^{M}$ are the stress-energy tensor of the matter fields
and an arbitrary 10-dimensional null vector on the background metric,
respectively.
From the Einstein equation $G_{MN}=8\pi T_{MN}$,
this condition is rewritten as 
\begin{align}\label{eq:NEC:0a}
  \cR^{\mathit{(Ein)}}_{MN}\xi^M\xi^N\ge 0\,,
\end{align}
where $\cR^{\mathit{(Ein)}}_{MN}$ is the Ricci tensor computed for the $10$d Einstein-frame metric \eqref{eq:metric:Ein}.
This condition finally leads to
\begin{subequations}
\label{eq:NEC:1}
\begin{align}
\label{eq:NEC:1a}
(\ti\alpha +\ti\beta -\delta) (4\ti\alpha -4\ti\beta +5\delta)
+4 r^{-2\delta} &\ge 0\,, \\
\label{eq:NEC:1b}
8\ti\alpha\ti\beta -5\delta (\ti\alpha +\ti\beta) +5\delta ^2&\le 0\,,
\end{align}
\end{subequations}
which must be satisfied everywhere.
The net condition of \eqref{eq:NEC:1a} is given at $r\to\infty$ for $\delta>0$ and $r\sim r_{\mathit{IR}}$ for $\delta < 0$.

\subsubsection{Locality: area law of entanglement entropy}
For local quantum field theories, the UV divergent part of the entanglement entropy for a spatial region $A$ is proportional to the area of the boundary $\partial A$.
This is called the {\it area law} of entanglement entropy.
We can make use of this property as a probe for the locality of the theory.

On the gravity side of the gauge/gravity correspondence, the leading part of entanglement entropy is computed 
by Ryu-Takayanagi formula~\cite{Ryu:2006bv,Ryu:2006ef}, as
\begin{align}
\label{eq:HEEformula}
  S_A = \min_{\gamma_A}\f{\mathrm{Area}(\gamma_A)}{4G_N}\,,
\end{align}
where $\gamma_A$ is a surface on the bulk reaching to the boundary,
where $\partial\gamma_A=\partial A$.
This area is computed in the Einstein-frame.
In this point of view, the area law states that the minimal surface $\gamma_A$ extends into the bulk, rather than sticks to the boundary (which leads to {\it volume law}).
By considering narrow stripes as $A$, this results in a simple condition,
\begin{align}
\label{eq:const:alpha+beta}
  \alpha + \beta > 0\,.
\end{align}

\subsection{Quark Masses}
\label{sec:QuarkMasses}
Because in this paper we are interested in emergent phenomena
appearing in heavy-quark limit,
we also need some constraints to realize such theory with heavy quarks.

The constituent quark mass $\ti{m}_Q$ is obtained as the mass of the
fundamental string connecting the horizon and the turning point $r=r_0$ of the D7-brane.
The mass $\ti{m}_Q$ is expressed as a function of $r_0$ and should increase as $r_0$ becomes large.
Since we can show that
\begin{align}\label{eq:tilmQ}
  \ti{m}_Q \sim r_0^{\alpha-\beta}\,
\end{align}
(where $r_0^0$ implies $\log{r_0}$)
from \eqref{eq:constraints:constituentmass},
we obtain the condition 
\begin{align}\label{eq:const:alpha-beta}
  \alpha-\beta\ge 0.
\end{align}

We also need to take the current quark mass $m_Q$ into account.
Because $m_Q$ can be regarded as the physical quark mass in weak coupling limit, 
it corresponds to the mass of the D3-D7 string on a flat background geometry.
Furthermore, it is in turn equal to the mass of the string connecting the D7-brane and the $\theta=\pi/2$ hypersurface, on the boundary $r\to\infty$.
We require that we can take this $m_Q$ to be very large but a finite (non-divergent) quantity,
together with a large value of $\ti{m}_Q$ at the same time.
This condition finally leads to quite strong constraints,
\begin{align}\label{eq:const:delta}
  \delta&=0\,,\\
\label{eq:const:gamma}
  \gamma&=-\frac{3}{4}\left(\alpha-\beta-\frac{1}{\alpha-\beta}\right) \,.
\end{align}

Under \eqref{eq:const:delta}, 
the rescaling symmetry \eqref{eq:rescaling:delta=0} 
implies that $\Theta(r)$ can be actually written as a function of $s$ \eqref{eq:s}.
Therefore, the expression for the constituent mass $m_Q$ \eqref{eq:constraints:currentmass} leads to
\begin{align}\label{eq:mQ}
  m_Q
  = \f{R}{2\pi\alpha'}\,r_0^{\alpha-\beta}\l\{\lim_{s\to\infty}s^{\alpha-\beta}\l(\f{\pi}{2}-\Theta(s)\r)\r\}
\propto r_0^{\alpha-\beta}\,,
\end{align}
which in turn shows $m_Q\sim \ti{m}_Q$.

\subsection{Yet Additional Constraints}
In addition to the constraints above,
there are also some conditions which we possibly should take into account.
We will not demand the conditions below at the first place,
and discuss them together with the results of the computations finally.

\subsubsection{Weakly coupled gravity}
Throughout this paper, we will rely on {\it classical} calculations
on the gravity side.
Then we have to ensure that 
the effective string coupling $e^\phi$ is small everywhere we work.
Since we will focus on the UV (large $r$) region 
where the dilaton behaves as \eqref{eq:dilaton}, we need
\begin{align}\label{eq:positivegamma}
  \gamma \ge 0\,,
\end{align}
otherwise we have to live along with horrible quantum gravity.
Together with the expression of $\gamma$ \eqref{eq:const:gamma},
we obtain 
\begin{align}\label{eq:const:alpha-beta:2}
  \alpha - \beta \le 1\,.
\end{align}

\subsubsection{Small curvature of bulk spacetime}
For the validity of the classical gravity,
the spacetime curvature has to be sufficiently small compared to the Planck scale.
In particular, we need that scalar components of the curvature do not diverge at $r\to\infty$.
By looking at the behaviors of $\cR$, $\cR_{MN}\cR^{MN}$ and $\cR_{MNPQ}\cR^{MNPQ}$,
we obtain a simple condition
\begin{align}
\label{eq:const:curv_cond}
  \beta \le 0\,,
\end{align}
under \eqref{eq:positivegamma}.
The details are given in Appendix~\ref{app:Curvature}.

\subsubsection{Proportionality of quark and meson masses}
\label{sec:mass_prop}
When the conventional quark model works well on the field theory side,
the leading parts of the meson masses are proportional to the quark mass.

As we will show in \S\ref{sec:hiddensym}, the meson masses behave as
\begin{align}
  M \propto r_0^{\alpha+\beta}\,,
\end{align}
as a function of $r_0$.
Comparing with \eqref{eq:mQ} (or \eqref{eq:tilmQ}), we get
\begin{align}
  M \propto (m_Q)^{\f{\alpha+\beta}{\alpha-\beta}}\,,
\end{align}
when $m_Q$ is large.
This implies that the meson mass is not proportional to the quark mass
when we take $\beta\ne 0$.
This phenomenon suggests the existence of a non-trivial strong-coupling effect, 
which is absent in standard interpretations of QCD.
It does not lead to any immediate inconsistency as a flavored gauge theory.
But if we do not like it, we can consider the additional constraint,
\begin{align}\label{eq:const:beta}
  \beta = 0\,.
\end{align}

\subsection{Possible Parameter Region}
\begin{figure}[htb]
   \centering
   \includegraphics[width=0.4\textwidth]{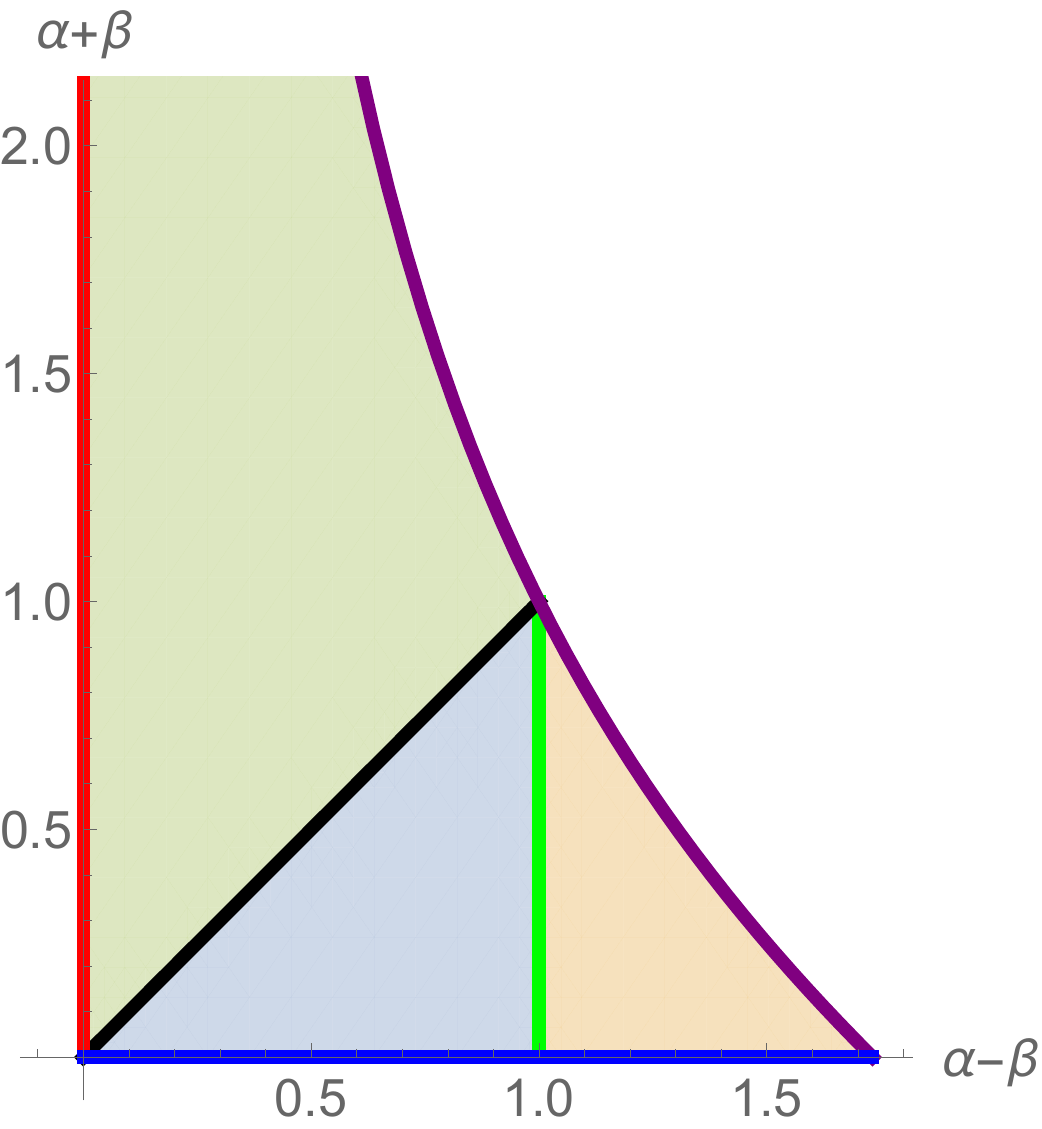}
  \caption{Constraints for the parameters $(\alpha,\beta)$.
  The bounds for \eqref{eq:const:alpha+beta}, \eqref{eq:const:alpha-beta}, \eqref{eq:NEC:2}, \eqref{eq:const:alpha-beta:2} and \eqref{eq:const:curv_cond}
  are displayed by the blue, red, purple, green and black lines respectively.
  They altogether yield the bluely painted triangle domain.
  }
   \label{fig:ConstraintsForParameters}
 \end{figure}

Under the conditions \eqref{eq:const:alpha+beta}, \eqref{eq:const:alpha-beta}, \eqref{eq:const:delta} and \eqref{eq:const:gamma},
the null energy condition \eqref{eq:NEC:1} simply becomes
\begin{align}\label{eq:NEC:2}
  \ti\beta \le 0
\quad\Leftrightarrow\quad
  \alpha + \beta \le \f{\zeta}{2}\,,
\end{align}
where we defined $\zeta$ as
\begin{align}\label{eq:zeta}
  \zeta \equiv 2(\alpha - \beta + 2\gamma) \;
\l(= \f{3}{\alpha-\beta} - (\alpha-\beta)\r)\,,
\end{align}
for later convenience.
This in turn automatically implies
\begin{align}\label{eq:const:alpha-beta:s3}
  \alpha - \beta \le \s{3}\,,
\end{align}
even without the condition \eqref{eq:positivegamma}.
Then the allowed possible parameter domain on $(\alpha-\beta,\alpha+\beta)$-plane
is shown graphically in \fgref{fig:ConstraintsForParameters}.

\section{Equations of Motion}
\label{ch:EOM}

In this section we consider the configuration and fluctuations 
of the D7-brane obeying the DBI action \eqref{eq:DBI}.
They lead to the equations which determine the meson spectra on the boundary theory.

We will focus on the case of 
$\delta=0$ \eqref{eq:const:delta} here.
We do not use the constraint for $\gamma$ \eqref{eq:const:gamma}
in this section 
unless specially noted,
because we used some results of this section in advance to derive \eqref{eq:const:gamma}.
Equations of motion for general $\delta$ are given in Appendix~\ref{sec:EoM:general}.

\subsection{Static Configuration of D7-Brane}
Here we derive the equations of motion for the D7 configuration and its fluctuations.
When $\delta=0$, we can combine
$r$ and $\Omega_5$ in \eqref{eq:metric:str}
as $dr^2+r^2d\Omega_5^2$, making a flat $\bR^6$.
We then divide this $\bR^6$ part into $\bR^4\times\bR^2$ as
\begin{align}
  dr^2 + r^2d\Omega_5^2 &= d\rho^2 + \rho^2d\Omega_3^2 + d\lambda^2 + \lambda^2d\psi^2\,.
\end{align}
At this time, the embedding profile of the D7-brane can be described by
$\lambda(\rho,\Omega_3,x^\mu)$ and $\psi(\rho,\Omega_3,x^\mu)$.
Compared to the coordinate \eqref{eq:S5coord}, we have the relation
\begin{align}
\label{eq:rholambda:transf}
  \rho = r\sin\theta\,,
\quad
  \lambda = r\cos\theta\,.
\end{align}

Now we can write the static embedding profile of the D7-brane as
\begin{align}\label{eq:L}
  \lambda &= L(\rho)\,,
\quad
  \psi = 0\,,
\end{align}
corresponding to \eqref{eq:setup:D7background}.
Then the D7 turning point is now $\rho=0$ and
the conditions \eqref{eq:Theta:smoothness} are translated to
\begin{align}\label{eq:L:initial}
  L(0) = r_0\,,
\qquad
  L'(0) = 0\,.
\end{align}
Substituting \eqref{eq:L} into the DBI action \eqref{eq:DBI},
the action for $L(\rho)$ is given as
\begin{align}\label{eq:SL}
  S_L
  &= -\f{2\pi^2T_7R^4 V_4}{g_0} 
\int_0^\infty\!\!d\rho\,\rho^3 \left(L^2+\rho ^2\right)^{2(\alpha - \beta +\gamma-1)}
\sqrt{\dot{L}^2+1}\,.
\end{align}
A dot (\,$\dot{}$\,) represents a derivative by $\rho$ hereafter,
unless otherwise noted.
$V_4$ is the 4D spacetime volume, $V_4=\int d^4x$.
This leads to the equation of motion
\begin{align}\label{eq:eom:L}
  \ddot{L} 
  + \f{3\dot{L}(1+\dot{L}^2)}{\rho} 
  + C\,\f{\rho\dot{L}-L}{\rho^2+L^2}(1+\dot{L}^2)
  &= 0\,,
\end{align}
where 
\begin{align}\label{eq:C}
  C &= 4(\alpha - \beta + \gamma - 1)\,.
\end{align}

\subsection{Fluctuation Modes}
On the static configuration $L(\rho)$ obeying \eqref{eq:eom:L},
we consider linear fluctuation modes
\begin{align}
  \delta\lambda \sim e^{-ik_{\mu}x^{\mu}}Y_{\ell mn}(\Omega_3)\Lambda(\rho) 
\quad\text{and}\quad
  \delta\psi \sim e^{-ik_{\mu}x^{\mu}}Y_{\ell mn}(\Omega_3)\Psi(\rho)\,,
\end{align}
where $Y_{\ell mn}(\Omega_3)$ are the 3-dimensional spherical harmonics. 
Furthermore, the linear fluctuation modes of the D7 world-volume gauge field
can be also written as
\begin{align}
  \delta A_x \sim e^{-ik_{\mu}x^{\mu}}Y_{\ell mn}(\Omega_3)a(\rho)\,.
\end{align}
In the viewpoint of holography, the eigenmodes of $\Lambda$, $\Psi$ and $a$
respectively correspond to scalar, pseudo-scalar and vector mesons 
on the boundary gauge theory.
Since we are interested in the degeneration of the spectrums of pseudo-scalar and vector mesons, 
we will focus on $\Psi$ and $a$ below.
For each modes labeled by $(k^\mu,\ell,m,n)$, 
the quadratic-order actions read
\begin{align}
  S_{a} &= -\f{4\pi^4\alpha'^2R^2T_7V_4}{g_0}\int_0^\infty\!\!d\rho\,
  \rho(\rho^2+L^2)^{2(-\beta+\gamma-1)}
  \bigg[
  \f{\rho^2(\rho^2+L^2)^{\alpha+\beta+1}}{\s{1+\dot{L}^2}}\dot{a}^2 
\nonumber\\&\hspace{4cm}+
  \s{1+\dot{L}^2}\l\{k_\mu k^{\mu}R^2\rho^2+\ell(\ell+2)(\rho^2+L^2)^{\alpha+\beta+1}\r\}a^2
\bigg]\,,\\
  S_{\Psi} &= -\f{\pi^2R^4T_7V_4}{g_0}\int_0^\infty\!\!d\rho\,
  \rho L^2(\rho^2+L^2)^{\alpha-3\beta+2\gamma-3}
  \bigg[
  \f{\rho^2(\rho^2+L^2)^{\alpha+\beta+1}}{\s{1+\dot{L}^2}}\dot{\Psi}^2
  \nonumber\\&\hspace{4cm}+
  \s{1+\dot{L}^2}\l\{k_\mu k^{\mu}R^2\rho^2+\ell(\ell+2)(\rho^2+L^2)^{\alpha+\beta+1}\r\}\Psi^2
\bigg]\,.
\end{align}
These actions are quite similar to each other,
and the only difference is the overall functional factor in the integrand
(i.e., Lagrangian).
They lead to the equations of motion for $a$ and $\Psi$ as
\begin{align}
\label{eq:eom:A}
  \ddot{a} + P_1(\rho)\dot{a} + P_0(\rho) a
  &= 0\,,\\
\label{eq:eom:psi}
  \ddot{\Psi} + \l[P_1(\rho) + Q_1(\rho)\r]\dot{\Psi} + P_0(\rho) \Psi
  &= 0\,,
\end{align}
where 
\begin{align}
\label{eq:a0}
  P_0(\rho) &= (1+\dot{L}^2)\l[\f{(-k_\mu k^\mu)R^2}{(\rho^2+L^2)^{\alpha+\beta+1}} - \f{\ell(\ell+2)}{\rho^2}\r]\,, \\
\label{eq:a1}
  P_1(\rho) &= 
  \f{3}{\rho} - \f{\dot{L}\ddot{L}}{1+\dot{L}^2}
  + 2(\alpha-\beta+2\gamma-1)\,\f{\rho+L\dot{L}}{\rho^2+L^2}
  \;\l(= \partial_\rho\log\l[\f{\rho^3(\rho^2+L^2)^{\alpha-\beta+2\gamma-1}}{\s{1+\dot{L}^2}}\r]\r)\,,
\\
\label{eq:b1}
  Q_1(\rho) &=
  \f{2\dot{L}}{L} + 2(\alpha-\beta-1)\,\f{\rho+L\dot{L}}{\rho^2+L^2}
  \;\l(= \partial_\rho\log\l[L^2(L^2+\rho^2)^{\alpha-\beta-1}\r]\r)\,.
\end{align}

In terms of holography, the mode functions of the 4D meson (quarkonium) excitations are given by the normalizable solutions to the bulk linear EoM's \eqref{eq:eom:A} or \eqref{eq:eom:psi}.
We will focus on the $s$-wave modes on the $S^3$, that is, $\ell=0$,
since we are not so interested in such \text{(R-)charged} mesons.
The 4D momentum square $(-k_\mu k^\mu)$ appearing in \eqref{eq:a0} represents the meson mass square, $M^2$.
For the mode functions we need to require the smoothness at $\rho=0$, i.e., $\dot{a}(0)=\dot{\Psi}(0)=0$,
as well as the normalizability at infinity ($\rho\to\infty$).
These two conditions cannot be satisfied at the same time in general,
and so we need to choose special values for $M^2$ for it.
This leads to discrete spectrums of the mesons, as is expected in flavors-confined phases of QCD-like theories.

As we can see from \eqref{eq:eom:A} and \eqref{eq:eom:psi}, 
the equations for $\Psi(\rho)$ and $a(\rho)$
are different by the term of $Q_1(\rho)\dot{\Psi}$.
Therefore, if $R_1(\rho)\equiv \sfrac{Q_1(\rho)}{P_1(\rho)}$ is sufficiently small all over the region of $\rho$, the degeneration of pseudo-scalar and vector mesons is automatically realized.
When $Q_1(\rho)$ vanishes all over the range of $\rho$, 
the whole spectra exactly agree with each other.
It occurs when
\begin{align}\label{eq:exactdegeneracy}
  \alpha-\beta-1=\gamma=0\,,
\end{align}
in which case 
\eqref{eq:eom:L} is solved as $L(\rho)=\text{const}$.
In fact, this is consistent with the constraint for $\gamma$ \eqref{eq:const:gamma} 
and so can be actually realized.
Note that \eqref{eq:exactdegeneracy} includes the case of supersymmetric $\mathrm{AdS}_5\times S^5$,
i.e., $(\alpha,\beta,\gamma)=(1,0,0)$,
and even more various geometries in addition to it.

In addition, we can rewrite the equations \eqref{eq:eom:A} and \eqref{eq:eom:psi} 
to the form of the standard eigenvalue problems of Schr\"odinger equations.
It is dealt with in Appendix~\ref{sec:Schroedinger},
although we will not directly use it in our computations later.

\subsection{Symmetry of the Equations of Motion and Mass Spectra}
\label{sec:hiddensym}
As we noted in \S\ref{sec:rescaling},
the rescaling transformation \eqref{eq:rescaling:delta=0}-\eqref{eq:A:redef} is a symmetry of 
the equations of motion \eqref{eq:eom:L}-\eqref{eq:b1}.
In terms of our current notations, it is translated as
\begin{align}
  L \to c\, L\,,
\quad
  \rho \to c\,\rho\,,
\quad
  k \to c^{\alpha+\beta}k\,,
\end{align}
which in turn imply
\begin{align}
  r_0\to c\,r_0\,,
\quad
  M \to c^{\alpha+\beta}M\,.
\end{align}
This means that the meson mass $M$ scales as
\begin{align}
  M \propto r_0^{\alpha+\beta}\,,
\end{align}
when we change the place of the turning point $r_0$.

Therefore the structure of the spectrum is invariant under the change of the quark mass.
It is not surprising, because 
we took the UV asymptotic form for the background geometry \eqref{eq:background}
from the beginning.
This suggests that the mass spectra we get from our model is 
those at heavy quark limit.

\subsection{Asymptotic Properties of the Equations of Motion}
\label{ch:asymptotics}
In this subsection,
we will see the asymptotic behavior of $L(\rho)$
and other functions in the small and large $\rho$ regions.

\subsubsection{Small $\rho$ behavior}
From \eqref{eq:L:initial},
we assume the leading behavior of $L(\rho)$ at $\rho\ll 1$ as
\begin{align}
  L(\rho) \simeq r_0 + q \rho^{n}\,,
\end{align}
where $n>1$ and $q$ are some constant.
Substituting it into \eqref{eq:eom:L}, the leading terms are
\begin{align}
   n(n+2)q\rho^{n-2} - \f{C}{r_0} \simeq 0\,.
\end{align}
Unless $C=0$, this equation can be satisfied at the leading order
only when the two terms are balanced, that is, $n=2$.
At that time this equation yields $q = \sfrac{C}{(8r_0)}$, then as a result we obtain
\begin{align}
\label{eq:smallrho:L}
  L(\rho) = r_0 + \f{C}{8r_0}\rho^2 + o(\rho^2)\,.
\end{align}
When $C=0$, we have an exact solution $L(\rho)\equiv r_0$,
which is included in \eqref{eq:smallrho:L} as a special case. 
From \eqref{eq:a0}, \eqref{eq:a1} and \eqref{eq:b1}, this leads to
\begin{align}
  P_0(\rho)\simeq M^2R^2{r_0}^{-2(\alpha+\beta+1)}\,,
\qquad
  P_1(\rho)\simeq \f{3}{\rho}\,,
\qquad
  Q_1(\rho)\simeq \f{C}{2r_0}(\alpha-\beta)\rho\,,
\end{align}
when $\ell=0$.
Therefore \eqref{eq:eom:A} and \eqref{eq:eom:psi} are always 
identical equations in this region ($\rho\ll 1$),
and the asymptotic solution for them is given by
\begin{align}
  A(\rho) \propto \psi(\rho) \propto 1 - \f{M^2R^2r_0^{-2(\alpha+\beta+1)}}{8}\rho^2 + o(\rho^2)\,.
\end{align}

\subsubsection{Large $\rho$ behavior}
When $C\ne 0$ and $\rho\gg R$, we consider several ansatz for 
the leading behavior of $L(\rho)$ as,
\begin{align}
  1.&~L(\rho) \simeq \ti{q}\rho^{\kappa}~(\kappa>1)\,,
&
  2.&~L(\rho) \simeq \ti{q}\rho\,,
\nonumber\\
  3.&~L(\rho) \simeq \ti{q}\rho^{\kappa}~(\kappa<1)\,,
&
  4.&~L(\rho) \simeq p + \ti{q}\rho^{\kappa}~(\kappa<1)\,,
\end{align}
which respectively give different leading forms for \eqref{eq:eom:L}.
Out of these, only $3.~L(\rho) \simeq \ti{q}\rho^{\kappa}\; (\kappa<1)$ has a solution, which is given by
\begin{align}
\label{eq:kappa}
  \kappa^2 + (C+2)\kappa - C =0
\quad\Leftrightarrow\quad
  \kappa = \kappa_{\pm} &\equiv \f{-(C+2)\pm\s{(C+4)^2-12}}{2}\,.
\end{align}
This solution again includes $L(\rho)\equiv r_0$ for $C=0$,
because $\kappa_+=0$ when $C=0$.

This asymptotic solution of $L(\rho)$ leads to the leading behavior of $P_0(\rho)$, $P_1(\rho)$ and $Q_1(\rho)$ as
\begin{align}
\label{eq:largerho:a0a1b1}
  P_0(\rho) &\simeq M^2R^2\rho^{-2(\alpha+\beta+1)}\,,
\qquad
  P_1(\rho) \simeq (\zeta + 1)\rho^{-1}\,,
\nonumber\\
  Q_1(\rho) &\simeq 2(\alpha - \beta + \kappa - 1)\rho^{-1}\,,
\end{align}
where
$\zeta$ is the same one as \eqref{eq:zeta}.
By using this asymptotics,
the asymptotic form of the normalizable solutions 
for \eqref{eq:eom:A} and \eqref{eq:eom:psi} are respectively given as
\begin{align}
\label{eq:largerho:a}
  a(\rho)&\sim \rho^{-\zeta}\,,\\
\label{eq:largerho:Psi}
  \Psi(\rho)&\sim \rho^{-(C+2\kappa_{+}+2)} = \rho^{-\s{(C+4)^2-12}}\,,
\end{align}
where we used the condition \eqref{eq:const:alpha+beta}.
In order for \eqref{eq:largerho:Psi},
we have to choose
\begin{align}\label{eq:L:largerho}
  L\sim \rho^{\kappa_+}\,,
\end{align}
out of $\kappa_{\pm}$ \eqref{eq:kappa},
otherwise we would have $\Psi\sim \rho^{\s{(C+4)^2-12}}$, which is not normalizable.

After we find \eqref{eq:const:gamma}, we can see that the asymptotics
\eqref{eq:largerho:a} and \eqref{eq:largerho:Psi}
are actually identical (i.e., $\zeta=\s{(C+4)^2-12}$).

\section{Numerical Computations}
\label{ch:Numerical_results}

\subsection{Method}
Now let us compute the numerical values of the meson masses from the equations of motion
\eqref{eq:eom:L}, \eqref{eq:eom:A} and \eqref{eq:eom:psi}.
Our basic strategy is the simple shooting methods.
First we temporary fix an arbitrary value for $M$,
and impose a boundary condition $A'(0)=0$ or $\psi'(0)=0$.
Then we solve the equation \eqref{eq:eom:A} or \eqref{eq:eom:psi}
from $\rho=0$ toward larger $\rho$,
by using the solution of \eqref{eq:eom:L}.
Then, for generic values of $M$, the solution for $A(\rho)$ or $\psi(\rho)$ goes to a nonzero constant at $\rho\to\infty$.
Only for special discrete values of $M$, it will decay at $\rho\to\infty$, satisfying the normalizability.
Those values of $M$ are nothing but the meson masses.
We call the $i$-th excited states of pseudo-scalar and vector mesons 
as $\eta_i$ and $\Upsilon_i$ ($i=0,1,2,\dots$), respectively,
and their masses are referred to by $M_\eta^{(i)}$ and $M_\Upsilon^{(i)}$.

Since the equations of motion becomes singular at $\rho=0$,
we introduce a cutoff $\rho=\rho_\epsilon$ where we impose initial conditions for \eqref{eq:eom:L}
by using the $\rho\to 0$ asymptotic form \eqref{eq:smallrho:L}.
We also introduce a boundary cutoff $\rho_{\infty}$,
where we determine the meson masses by $A(\rho_\infty)=0$ or $\psi(\rho_\infty)=0$.
The authors implemented this procedure on a {\it Mathematica}\texttrademark\@  program
and carried out the series of numerical computations.
They adopted $\rho_\epsilon=10^{-25}$ and $\rho_\infty=2^{20}$,
and used 50-digits floating-point numbers.

\begin{figure}[htb]
  \centering
  \begin{tabular}{clccl}
\includegraphics[height=0.4\textheight]{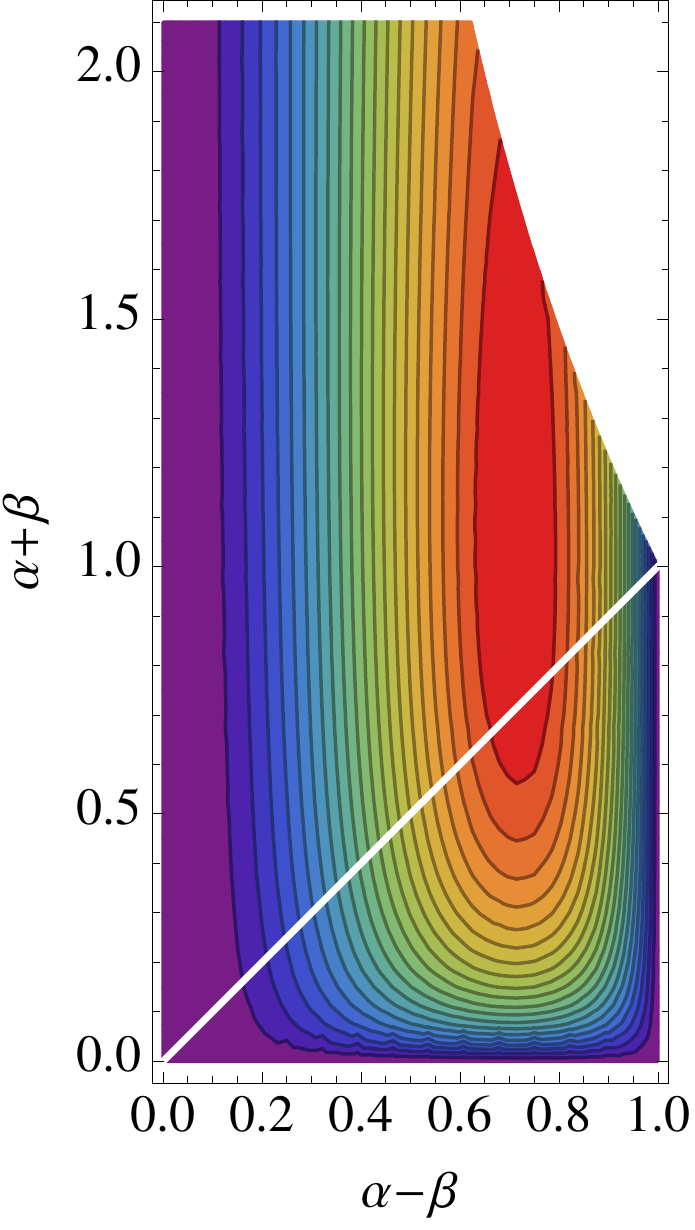}
&
\raisebox{0.035\textheight}[0pt][0pt]{
\includegraphics[height=0.25\textheight]{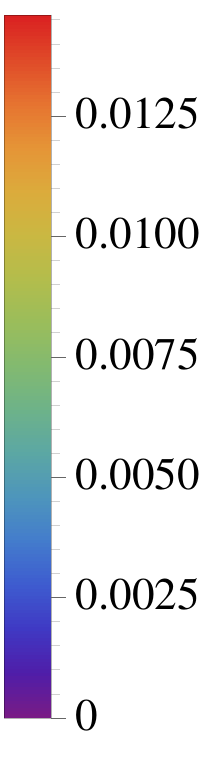}
}
&
\hspace{5mm}
&
\includegraphics[height=0.22\textheight]{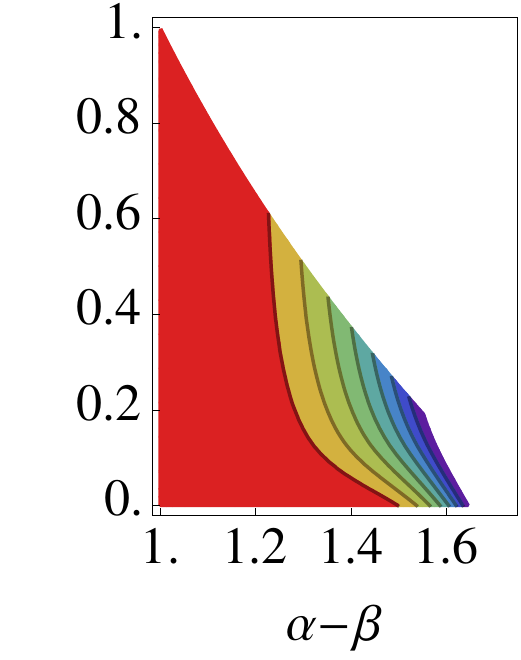}
&
\raisebox{0.035\textheight}[0pt][0pt]{
\includegraphics[height=0.25\textheight]{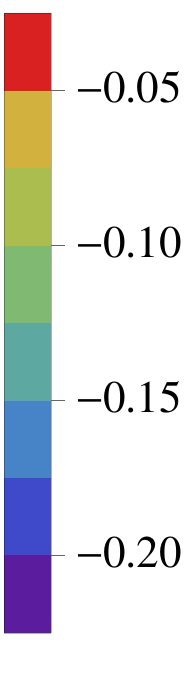}
}
\\
\multicolumn{2}{c}{(a)}
&&
\multicolumn{2}{c}{(b)}
   \end{tabular}
  \caption{Contour plots of the mass difference $(M_\Upsilon^{(0)}-M_\eta^{(0)})$ in the unit of $M_\eta^{(0)}=1$, 
on the $(\alpha-\beta,\alpha+\beta)$-plane
for {\bf (a)} $0\le \alpha-\beta\le 1$ and {\bf (b)} $1\le\alpha-\beta\le\s{3}$.
The white line in (a) shows $\beta=0$.
}
\label{fig:M0contour}
\end{figure}
\begin{figure}[htb]
  \centering
\includegraphics[width=0.6\textwidth]{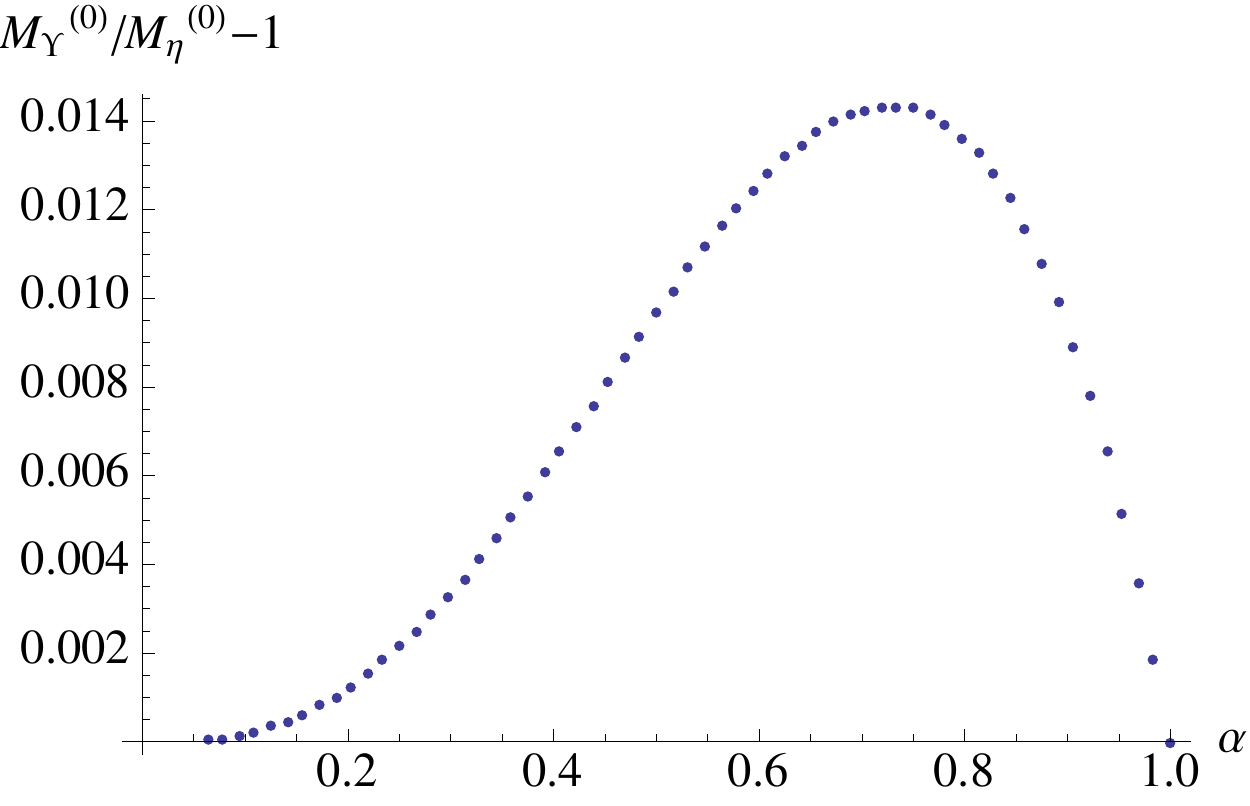}
  \caption{Plot of $M_\Upsilon^{(0)}-M_\eta^{(0)}$ on $\beta=0$ (white line on \fgref{fig:M0contour}(a)).}
  \label{fig:M0beta0}
\end{figure}
\subsection{Results for Ground State Masses}
The numerical plots for the meson mass difference $M_\Upsilon^{(0)}-M_\eta^{(0)}$ is shown in \fgref{fig:M0contour}.
In the region (a) $\alpha-\beta\le 1$ (corresponding to $\gamma\ge 0$),
$M_\Upsilon^{(0)}$ is slightly larger than $M_\eta^{(0)}$,
in agreement with the tendency observed in experiments 
for charmonia ($\eta_c(2980)$, $J/\psi(3097)$) and bottomonia ($\eta_b(9390)$, $\Upsilon(9460)$).
The mass difference is nonzero in general, but rather small everywhere
(up to $\simeq 1.5\%$).
We can observe that exact degeneration is realized on $\alpha+\beta\to 0$ or
$\alpha-\beta\to 0$ limits respectively,
as well as on $\alpha-\beta=1$ \eqref{eq:exactdegeneracy}.
In particular, the $\beta=0$ cross section is shown in \fgref{fig:M0beta0}.
The exact degeneration at $(\alpha,\beta)=(1,0)$ on this graph is the trivial one caused by the supersymmetry.
It is remarkable that degeneration also takes place in the opposite limit $\alpha\to 0$.

On the contrary, in the region (b) $\alpha-\beta>1$,
the order of $M_\Upsilon^{(0)}$ and $M_\eta^{(0)}$ is reversed,
and the mass splitting becomes relatively large (more than 30\% at most).
Since this region breaks the weak gravity condition \eqref{eq:const:gamma},
our classical calculation is not reliable in general,
due to quantum-gravitational effects.

In summary, we conclude that the heavy quark spin symmetry is 
realized for the lowest quarkonia,
not exactly but in a high accuracy,
in our generic parameter space of the dual geometry
at least where we can rely on the classical gravity.

\subsection{Results for Excited State Masses}
In this section, the excited states of the quarkonia are investigated.
We compare the results of $(\alpha,\beta)=(1,0)$ and
$(\alpha,\beta)=(1/16,0)$.
The result of $(\alpha,\beta)=(1,0)$ holds the supersymmetry.
On the other hand, the supersymmetry is broken in the result of
$(\alpha,\beta)=(1/16,0)$, but the vector and pseudo-scalar mesons are
almost degenerate in the ground states as seen in the previous section.
In \fgref{fig:M0beta0}, the mass difference in the ground state
decreases as $\alpha$ comes close to zero.
However, the value for $(\alpha,\beta)=(0,0)$ has singularity.
Instead of this, 
we will show the results for $(\alpha,\beta)=(1/16,0)$ which gives the smallest
mass difference in the small $\alpha$ region in our numerical
calculation.

\begin{figure}
 \begin{center}
  \includegraphics[width=0.6\textwidth,clip]{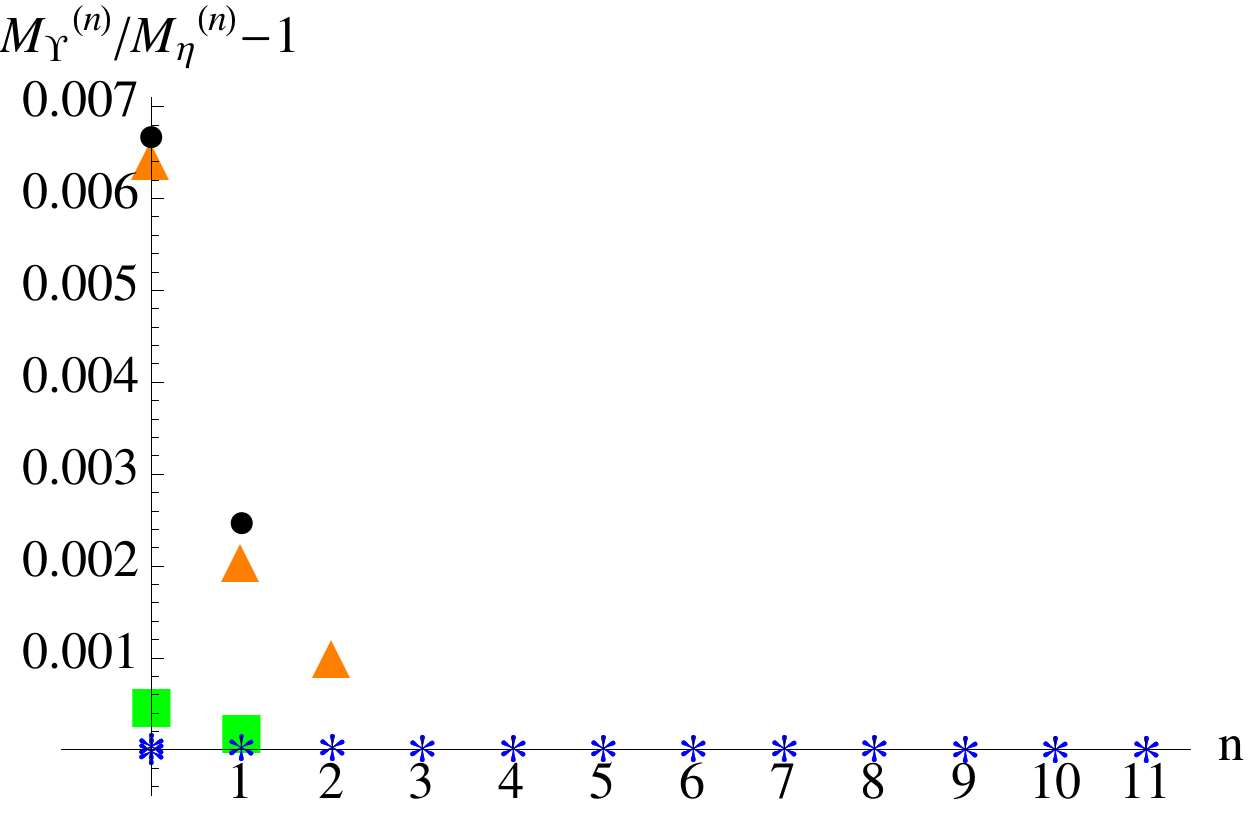}
  \caption{Plot of mass difference $M_{\Upsilon}/M_{\eta}-1$.
  Masses are measured from the ground state mass of the pseudo-scalar
  meson $M^{(1)}_{\eta}$.
  Asterisk symbol shows the results for $(\alpha,\beta)=(1/16,0)$.
  Dot is the experimental values for the bottomonium~\cite{Agashe:2014kda}.
  Triangle and square are the spectra of bottomonium and topponium,
  respectively, obtained by the quark model
  calculation in Ref.~\cite{Godfrey:1985xj}.
  }
  \label{fig:massdif_excited}
 \end{center}
\end{figure}

In \fgref{fig:massdif_excited},
the mass differences between the excited vector and pseudo-scalar mesons
with 
$(\alpha,\beta)=(1/16,0)$ are shown.
For $(\alpha,\beta)=(1,0)$ with the supersymmetry, 
the mass differences are exactly zero even in the excited states.
On the other hand, the results for $(\alpha,\beta)=(1/16,0)$ are not
zero but much smaller than
the experimental results for the bottomonia~\cite{Agashe:2014kda}
and also than the quark model
predictions~\cite{Godfrey:1985xj} for the bottomonia and topponia ($t\bar{t}$\,)\footnote{
Topponia are hypothetical states 
because top itself is too unstable due to the electroweak interaction.
However, they can be obtained in the
theoretical studies and 
are useful to see the quark mass dependence of meson spectra.
}. 
The mass differences $M_{\Upsilon}/M_{\eta}-1$ for
$(\alpha,\beta)=(1/16,0)$ are $4.7\times 10^{-3}\%$ for $n=0$ and 
$4.1\times 10^{-3}\%$ for $n=1$, while 
ones for the topponia in \fgref{fig:massdif_excited} are
$4.4\times 10^{-2}\%$ for $n=0$ and $1.4\times 10^{-2}\%$ for $n=1$.
As seen in the ground state, 
the small mass differences between pseudo-scalar and vector
mesons are also found in the excited states for $\alpha-\beta\leq 1$.

\begin{figure}
 \begin{center}
  \includegraphics[width=0.6\textwidth,clip]{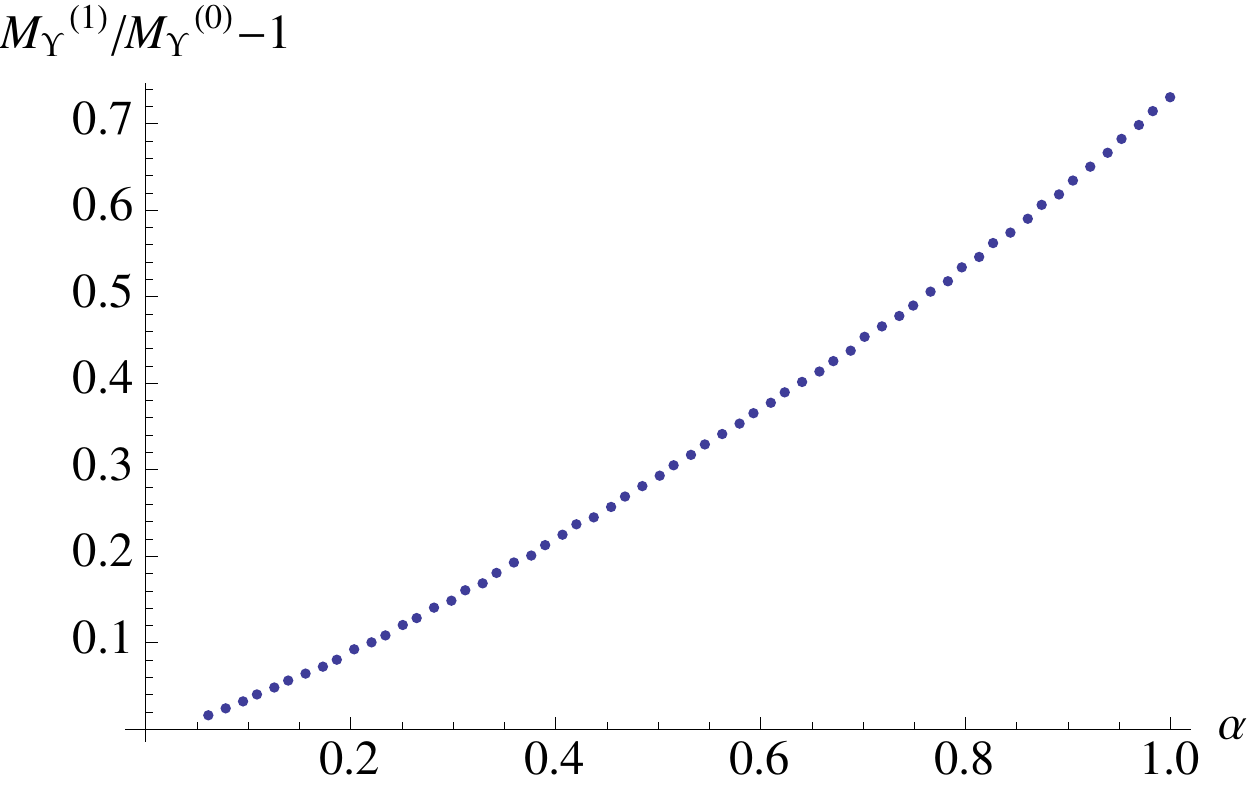}
  \caption{Plot of the $\alpha$ dependence of the mass difference
  between the first-excited and ground states in $\alpha< 1$ and $\beta=0$.}
  \label{fig:excited_energy_n0n1}
 \end{center}
\end{figure}

We estimate the excitation energies of the quarkonium spectra.
In \fgref{fig:excited_energy_n0n1},
the $\alpha$ dependence of the mass difference between the first-excited and
ground states of the vector meson, $M^{(1)}_{\Upsilon}/M^{(0)}_{\Upsilon}-1$, is shown in $\alpha\leq 1$ and
$\beta=0$.
The mass difference decreases with
decrease in $\alpha$.
The excitation energies for $(\alpha,\beta)=(1,0)$ 
are much larger than the ones in small $\alpha$ region,
while the mass degeneracies are realized in both cases.
Similar behavior is also obtained in the result of the pseudo-scalar mesons.

\begin{figure}
 \begin{center}
  \includegraphics[width=0.63\textwidth,clip]{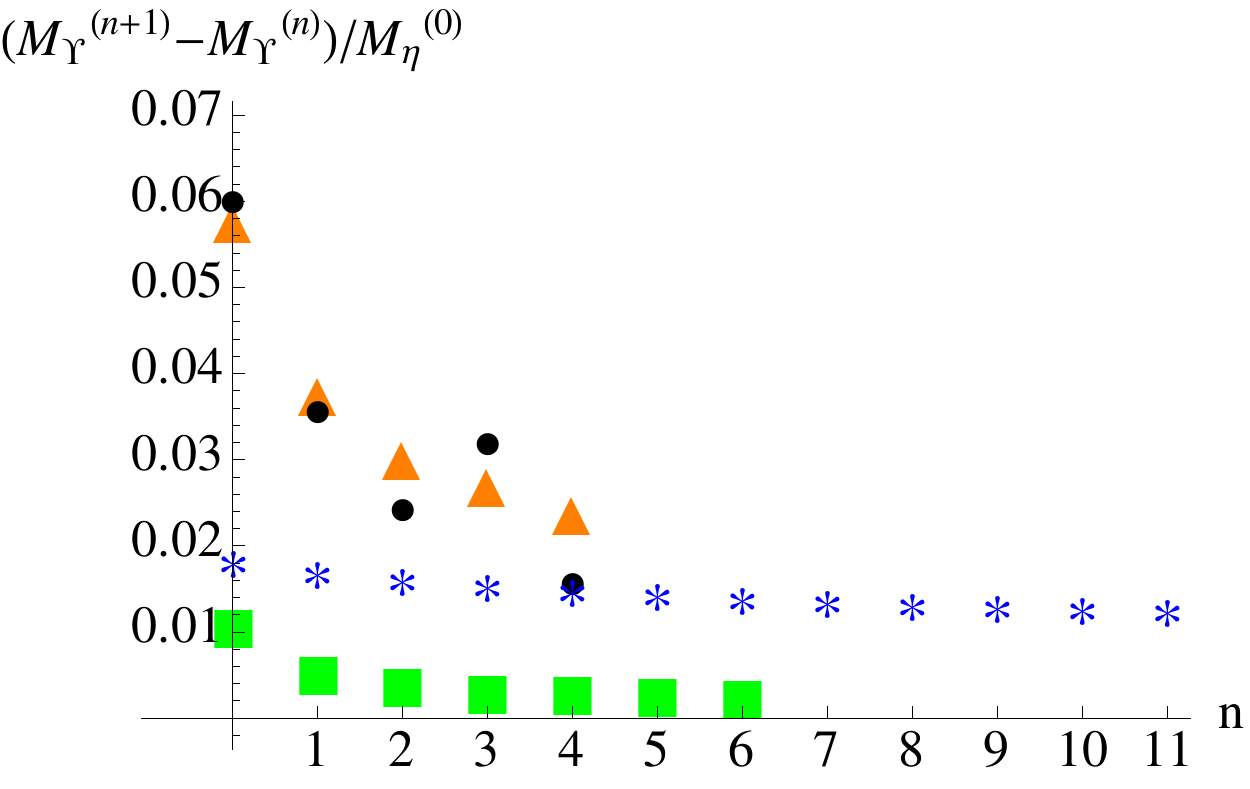}
    \caption{excitation energies for vector mesons, 
    $M^{(n+1)}_{\Upsilon}-M^{(n)}_{\Upsilon}$ for $n\geq 0$.
  The masses are normalized by $M^{(0)}_{\eta}$.
    Asterisk shows the results for $(\alpha,\beta)=(1/16,0)$.
    Dot is the bottomonium spectra in the experimental data.
    Triangle and square are the quark model predictions for bottomonia and
    topponia, respectively~\cite{Godfrey:1985xj}.
  }
  \label{fig:excited_energy_EXPvsHQS}
 \end{center}
\end{figure}

In \fgref{fig:excited_energy_EXPvsHQS},
excitation energies for $(\alpha,\beta)=(1/16,0)$ 
are shown as
$M^{(n+1)}_{\Upsilon}-M^{(n)}_{\Upsilon}$ for $n\geq 0$.
The masses are normalized by $M^{(0)}_{\eta}$.
In comparison with the experimental values and other theoretical studies,
the energies for $(\alpha,\beta)=(1/16,0)$ are comparable with those values as
seen in \fgref{fig:excited_energy_EXPvsHQS}.
On the other hand, the excitation energies for $(\alpha,\beta)=(1,0)$ are very large.
In \fgref{fig:excited_energy_EXPvsHQS},
the excitation energies 
go down slightly as $n$ increases.
The experimental and theoretical
results~\cite{Agashe:2014kda,Brambilla:2010cs,Godfrey:1985xj,Barnes:2005pb,Liu:2012ze,Aoki:2013ldr} also
show such tendency.
In the quark model calculation,
this behavior is understood to be caused by the linear component in the
confinement potential.

\section{Summary and Discussions}
\label{ch:Disscussion_Conclusion}

\paragraph{Summary}
In this paper, 
we have studied holography on
general D3-like gravitational backgrounds, with a flavor D7 brane.
We analyzed the conditions for the dual boundary theory being a physical QCD-like theory
--- stability,
locality,
current/constituent quark masses,
and some supplementary constraints
---
and determined the proper parameter region.
Finally, by solving the equations of motion of the DBI action of the D7 brane on such backgrounds,
we calculated the mass degeneracy of pseudo-scalar and vector quarkonia
of the boundary theory at heavy quark limit.
As a result, we found that the degeneracy is generically realized in a good accuracy, while there are tiny mass splittings.

\paragraph{Interpretation of our results}
For the most part, 
our results show good qualitative agreements 
with 
those of phenomenological or perturbative QCD effective models,
as well as experiments.
It suggests a universality of (approximate) heavy quark spin symmetry, 
even for generic, strongly coupled theories.

Besides it,
our new important observation is the existence of the small breaking of the heavy quark symmetry,
which was absent in those effective models.
This difference is quite interesting, 
because our holographic calculation includes non-perturbative effects
as well as the perturbative contributions of the gauge theory.
Compared to the perturbative approaches, that non-perturbativity explains the disagreement.
It also suggests that the phenomenological quark model should include 
small spin-dependent terms in the quark-antiquark potential function,
coming from a similar origin.
Conversely speaking,
our results imply that such non-perturbative effects are not too large,
and the conventional assumptions of those effective models are good approximations
to some extent.
We expect that lattice simulations will finally confirm our predictions, 
although they are still suffering from too large computational cost,
especially for heavy quark theories.

Although our expectation is as above, 
perhaps a supporter for exact heavy quark symmetry
could defend it with some other logical possibilities below,
which could not be rejected immediately.

One would be a claim that we must tune the parameters to realize
the exact heavy quark symmetry
as well as the quark-meson mass proportionality \eqref{eq:const:beta},
for the boundary theory being a physical QCD-like one.
According to it, the only (asymptotic) physical point in our parameter space would be
$(\alpha,\beta,\gamma)=(0,0,\infty)$,
where the power-law ansatz \eqref{eq:background} breaks down and
we would have to replace it by different ones including $\exp$, $\log$, etc.
Holography on such geometry behaves slightly differently from standard ones \cite{Ogawa:logpreparation},
and such subtlety might be important for heavy quark physics.

Another, maybe relatively more presumable one is as follows.
Even when we are taking the heavy quark limit,
it is only after the limits of
large $N_c$ and strong ('tHooft) coupling
for classical gravity approximation.
This might suggest that
we cannot arrive at the regime where the heavy quark effect could beat those of large $N_c$ and strong coupling.
In other words, 
on the gravity side,
quantum or string corrections might work to cancel the mass splittings 
and to restore the heavy quark symmetry.
This discussion could be regarded to be
consistent with the relatively large mass splittings for $\gamma<0$ (\fgref{fig:M0contour}~(b)),
where quantum gravitational corrections are expected to be especially important.

\paragraph{Applications to supergravity solutions}
It is possible to perform the same analysis as this paper for supergravity background 
which has a known dual gauge theory, in particular. 
That would enable us to check the heavy quark symmetry in 
some non-trivial, concrete theories.
The $\cN=2$ examples suffer from the automatic mass degeneracy 
due to the supersymmetry, 
as we mentioned in the introduction. 
Any vector meson should be accompanied with a complex scalar mesons (a scalar plus a pseudo-scalar), 
as a consequence of the $\cN=2$ supermultiplet.%
\footnote{Even if the supersymmetry is broken at some scale, 
typically the supersymmetry may be restored at the heavy quark limit. 
For example, the famous example of the D4-D6 model with
broken supersymmetries by the anti-periodic boundary condition for fermions 
in the Kaluza-Klein circle \cite{Kruczenski:2003uq}, 
the spectra of mesons were analyzed in \cite{Jo:2011xq} which shows the mass degeneracy. 
However this is due to the supersymmetry restoration at the heavy quark limit.} 

So we need to consider $\cN=1$ or non-supersymmetric examples. 
The $\cN=1$ examples include Klebanov-Witten geometry \cite{Klebanov:1998hh} 
and its generalization such as 
supergravity on Sasaki-Einstein manifolds (see for example \cite{Martelli:2004wu}), 
and non-conformal theories such as 
gravity dual to $\cN=1$ super QCD where explicit gravity background solutions are known (see for example \cite{Casero:2006pt}). 
A peculiar $\cN=0$ model is a gravity on $\mathrm{AdS}_5\times S^5/\Gamma$ 
where $\Gamma$ is a discrete group \cite{Kachru:1998ys}. 
Depending on how the group $\Gamma$ acts on the sphere, 
possible flavor D-brane configurations are classified. 
When the D7-brane configuration is consistent with $\Gamma$,
as is the case for our computations in this paper,
the meson spectrum on the D7 is left intact 
and the (supersymmetric) mass degeneracy remains,
at the classical level on the gravity side.
At that time, however,
new twisted sectors arise, 
and they would break the supersymmetry and the mass degeneracy
through quantum ($1/N_c$-order) corrections.
Furthermore, applications to non-asymptotically-AdS supergravity solutions such as \cite{Klebanov:2000hb}
would be even more interesting.

\paragraph{Relation to T-duality?}
We point out that the heavy quark symmetry may originate in string T-duality. 
Any D-brane has scalar and vector excitations as its massless part of string fluctuations, and they are related by the T-duality. 
The T-duality is manifest in a flat background geometry. 
But in holography, the flavor D-brane is put in a curved spacetime, 
so in an appropriate limit, the T-duality symmetry may recover. 
It might be a plausible guess that the recovery of the T-duality may be related to the heavy quark limit 
at which the flavor D-brane is pushed toward the boundary of the geometry
where the background geometric structure simplifies. 
The role of supersymmetry may be important there, 
but generically the T-duality works even in the absence of the supersymmetry, 
as is obvious from D-branes in bosonic string examples. 
It would be interesting to pursue this direction
to find possible intrinsic and geometric origin of the holographic heavy quark symmetry.

\section*{Acknowledgements}
The authors thank
Norihiro Iizuka,
Carlos N\'u\~nez
and 
Koichi Yazaki
for fruitful discussions.
The work of N.O. is supported by the Special Postdoctoral
Researcher (SPDR) Program of RIKEN.
This work is partly supported by the interdisciplinary Theoretical Science (iTHES) Project of RIKEN.
The authors are also grateful to the anonymous referee of JHEP
for valuable suggestions to improve this paper.

\appendix
\section{Details of Derivation of Constraints for Parameters}
\label{ch:constraints}
In this Appendix, 
we give the details of the derivations of the various 
constraints we gave in \S\ref{ch:Constraints_para}
for the parameters
$(\alpha,\beta,\gamma,\delta)$ in the action~\eqref{eq:metric:str}.

\subsection{Null Energy Condition}
Here let us solve the null energy condition combined with Einstein equation,
\eqref{eq:NEC:0a},
\begin{align}
\label{eq:nec}
  \cR_{MN}^{(\mathit{Ein})}\xi^M\xi^N\ge 0\,.
\end{align}
The Ricci tensor $\cR_{MN}^{(\mathit{Ein})}$ is calculated from the Einstein-frame metric \eqref{eq:metric:Ein} as
\begin{align}
  \cR_{MN}^{(\mathit{Ein})}dx^{M}dx^{N} 
  = &-\f{\alpha(4\ti\alpha-4\ti\beta+5\delta)}{R^2}\,r^{2(\ti\alpha+\ti\beta)}\eta_{\mu\nu}dx^{\mu}dx^{\nu}
  + \l[-4\ti\alpha(\ti\alpha+\ti\beta)+5\delta(\ti\beta-\ti\delta)\r]\f{dr^2}{r^2}
\nonumber\\
  &+ \l[(4\ti\alpha-4\ti\beta+5\delta)(\ti\beta-\delta)r^{2\delta}+4\r]d\Omega_5^2\,.
\end{align}
The null vector can be given as
\begin{align}
  \xi_1^M\partial_M &= \partial_t + \partial_x\,,
\quad\text{or}\quad
  \xi_2^M\partial_M = 
  r^{-\ti\alpha}\partial_t 
  + (\cos\omega) \f{r^{\ti\beta+1}}{R}\partial_r 
  + (\sin\omega) \f{r^{\ti\beta-\delta}}{R}\partial_{\theta}\,, 
\end{align}
without loss of generality,
by the $\SO(1,4)\times\SO(6)$ isometry of \eqref{eq:metric:str}.
They lead to $R_{MN}\xi_1^M\xi_1^N\equiv 0$ and
\begin{align}
  \cR_{MN}^{(\mathit{Ein})}\xi_2^M\xi_2^N
  =
  \frac{r^{2\ti\beta}}{R^2}
  \Big[
  &2r^{-2\delta}
  + \l(2\ti\alpha ^2- 4\ti\alpha  \ti\beta + 3\ti\alpha\delta - 2\ti\beta^2 +7\ti\beta\delta -5\delta ^2\r)
\nonumber\\
  &+2\cos(2\omega) \l\{r^{-2\delta} -\l(-\ti\alpha ^2 -2\ti\alpha\ti\beta +\ti\alpha\delta +\ti\beta^2 -\ti\beta\delta \r)\r\}
  \Big]\,,
\end{align}
which gives 
\begin{subequations}
\label{eq:nec:result}
\begin{align}
\label{eq:nec:result:1}
  (\ti\alpha +\ti\beta -\delta) (4\ti\alpha -4\ti\beta +5\delta) +4 r^{-2\delta} &\ge 0\,,\\
\label{eq:nec:result:2}
  8\ti\alpha\ti\beta -5\delta (\ti\alpha +\ti\beta) +5\delta ^2 &\le 0\,,
\end{align}
\end{subequations}
from \eqref{eq:nec},
at $\omega=0$ and $\omega=\sfrac{\pi}{4}$ respectively.
Depending on the sign of $\delta$, they are furthermore rewritten as follows.
\paragraph{Case 1: $\bm{\delta=0}$}
When $\delta=0$, \eqref{eq:nec:result} becomes
\begin{subequations}
\label{eq:nec:result:delta=0}
\begin{align}
\label{eq:nec:result:1:delta=0}
  \hat\alpha^2 - \hat\beta^2 + 1&\ge 0\,,\\
\label{eq:nec:result:2:delta=0}
  \hat\alpha\hat\beta &\le 0\,.
\end{align}
\end{subequations}

\paragraph{Case 2: $\bm{\delta>0}$}
When $\delta>0$, \eqref{eq:nec:result:1} gives the most strict condition
at $r\to\infty$, as
\begin{subequations}
\label{eq:nec:result:delta>0}
\begin{align}
\label{eq:nec:result:1:delta>0}
  (\hat\alpha +\hat\beta -1) \l(\hat\alpha -\hat\beta +\f{5}{4}\r) &\ge 0\,,
\end{align}
while \eqref{eq:nec:result:2} becomes
\begin{align}
\label{eq:nec:result:2:delta>0}
  8\hat\alpha\hat\beta -5(\hat\alpha +\hat\beta) +5 &\le 0\,.
\end{align}
\end{subequations}

\paragraph{Case 3: $\bm{\delta<0}$}
When $\delta<0$, \eqref{eq:nec:result:1} is more strict when $r$ is smaller, 
but there is a bound $r\simeq r_{\mathit{IR}}$, where the UV form of the geometry \eqref{eq:background} breaks down.
Then the condition is
\begin{subequations}
\label{eq:nec:result:delta<0}
\begin{align}
\label{eq:nec:result:1:delta<0}
  (\hat\alpha +\hat\beta +1) \l(\hat\alpha -\hat\beta -\f{5}{4}\r) &\ge 
  -\f{r_{\mathit{IR}}^{-2\delta}}{\delta^2}\,,
\end{align}
while \eqref{eq:nec:result:2} becomes
\begin{align}
\label{eq:nec:result:2:delta<0}
  8\hat\alpha\hat\beta +5(\hat\alpha +\hat\beta) +5 &\le 0\,.
\end{align}
\end{subequations}

\subsection{Entanglement Entropy}
Since we are interested in QCD-like theories, 
the gluon sector consists of local degrees of freedom
and deconfined in UV regime.
The behavior of entanglement entropy is restricted by such conditions.
The UV divergent term should obey the so-called area-law, from the locality
of the theory.
We can deal with this condition on the gravity side 
by using Ryu-Takayanagi formula \eqref{eq:HEEformula}.

\begin{figure}
  \centering
  \includegraphics[height=0.2\textheight]{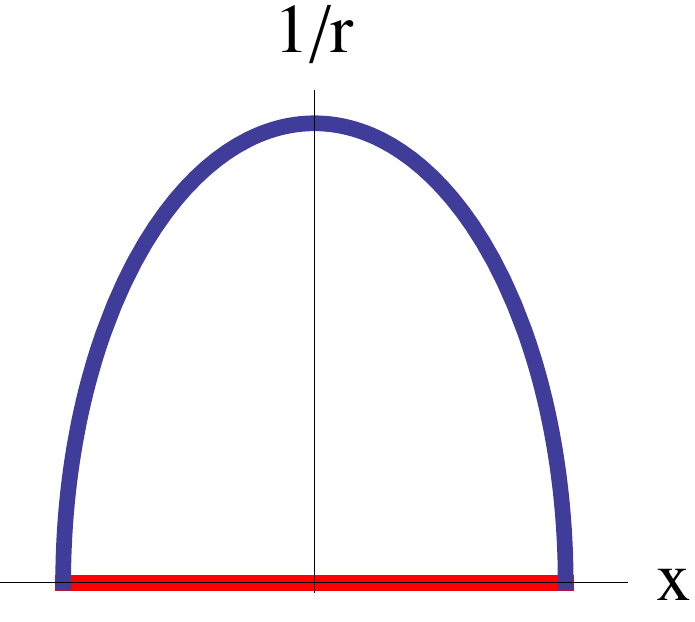}
  \caption{Candidates for the minimal surface profile. 
Blue line gives {\it area-law}, 
whereas red line leads to {\it volume law} 
for the UV divergent part of the entanglement entropy.}
  \label{fig:HEE}
\end{figure}

Let us take the subsystem $A$ to be a stripe with its width $\ell$, 
say $-\ell/2 < x < \ell/2$.
Then the corresponding minimal surface $\gamma_A$ should be given
by a curved profile $r=r(x)$, s.t., 
$r(\pm\ell/2)=r_{\infty}(\to\infty)$ and $r(0)=r_*$ which is the turning point of the surface.
It is drawn by the blue line in \fgref{fig:HEE},
on the $(x,1/r)$-plane.
Then the area is written down as
\begin{align}
  \mathrm{Area}(\gamma_A)
  &\propto A_2(\partial A)\int_{r_*}^{r_\infty}\!dr\,r^{3\ti{\alpha}-5\ti{\beta}+5\delta}\s{\l(\f{dx}{dr}\r)^2+r^{-2(\ti\alpha+\ti\beta+1)}}\,,
\end{align}
where $A_2(\partial A)$ is the area of $\partial A$ (which is divergent).
This is extremized by the solution of the Euler-Lagrange equation,
\begin{align}
\label{eq:HEE:profile}
  \f{dx}{dr} = \f{f(r_*)\,r^{-(\ti\alpha+\ti\beta+1)}}{\s{f(r)^2-f(r_*)^2}}\,,
\qquad
  \l(f(r)\equiv r^{3\ti\alpha-5\ti\beta+5\delta}\r)\,,
\end{align}
giving the UV divergent term of the area as
\begin{align}
\label{eq:HEE:UVarealaw}
  \mathrm{Area}(\gamma_A)_{UV} \sim A_2(\partial A)\,r_{\infty}^{\;\;2\ti\alpha-6\ti\beta+5\delta}\,.
\end{align}
The reality of the solution \eqref{eq:HEE:profile} requires
\begin{align}\label{eq:HEE:additional}
  3\ti\alpha - 5\ti\beta + 5\delta > 0\,
\quad&\Leftrightarrow\quad
  3\alpha - 5\beta + 5\delta + 8\gamma > 0\,
\nonumber\\
\quad&\Leftrightarrow\quad
  3\hat\alpha - 5\hat\beta > -5\,\mathrm{sign}(\delta)\,.
\end{align}
In order that this solution gives the true minimal area,
we need that it is smaller than the other candidate,
$r(x)\equiv r_{\infty}$,
shown by the red line in \fgref{fig:HEE}.
This second profile gives a ``volume law'' UV behavior,
which is computed as
\begin{align}
\label{eq:HEE:UVvolumelaw}
  \mathrm{Area}(\gamma_A)_{UV}^{\mathrm{vol\hyphen law}} \sim V_3(A)\,r_{\infty}^{\;\;3\ti\alpha-5\ti\beta+5\delta}\,,
\end{align}
where $V_3(A)=A_2(\partial A)\times\ell$.
Comparing \eqref{eq:HEE:UVarealaw} with \eqref{eq:HEE:UVvolumelaw},
we obtain
\begin{align}
\label{eq:HEE:alpha+beta}
  \ti\alpha + \ti\beta > 0
\quad\Leftrightarrow\quad
\alpha + \beta > 0\,.
\end{align}

Note that,
under \eqref{eq:const:delta} and \eqref{eq:const:gamma},
\eqref{eq:HEE:additional} will be equivalent to
\begin{align}
  \alpha + \beta < 2\zeta\,,
\end{align}
which 
will not affect the final conclusion of the physical parameter region.

\paragraph{Wilson loop}
Under the assumption that the background fluxes do not make
any important contributions,
we can also calculate the expectation values of the stripe-like
Wilson loops
which correspond to the quark-antiquark potential.
Unlike the entanglement entropy above, 
we deal with (1+1)-dimensional string world-sheets
and work on the string-frame metric \eqref{eq:metric:str}.
It can be written down as
\begin{align}
  -\log\l<\cW(\cC)\r>
  &\propto \min_{x(r)}\int_{r_*}^{r_\infty}\!dr\,r^{2\alpha}\s{\l(\f{dx}{dr}\r)^2+r^{-2(\alpha+\beta+1)}}\,.
\end{align}
We expect that it obeys the so-called perimeter-law
(which is essentially the same one as the ``area-law'' for entanglement entropy above),
in order that the system is a local gauge field theory
in a deconfined phase at short distances.
It is straightforward to follow the same prescription above,
and we find the conditions
\begin{align}
  \alpha > 0\,,
\qquad
  \alpha + \beta > 0\,.
\end{align}
They are always ensured by 
\eqref{eq:HEE:alpha+beta} and \eqref{eq:constraint:alpha-beta}.

Although the constraints from the entanglement entropy and the Wilson loop
may look almost equivalent,
that of the entanglement entropy is more rigorous 
because it depends only on the Einstein-frame metric \eqref{eq:metric:Ein}
and does not need any assumption for background fluxes.

\subsection{Conditions for Quark Masses}
\label{ch:constraints:quarkmass}
\subsubsection{Constituent quark mass}
\label{ch:constraints:constituentmass}
The constituent quark mass is given by the mass of the fundamental string
which connects the horizon and the turning point $r=r_0$ of the D7, as
\begin{align}\label{eq:constraints:constituentmass}
  \ti{m}_{Q} &= \f{1}{2\pi\alpha'}\int_{0}^{r_0}\!\s{-g_{tt}g_{rr}}\,dr
 = \ti{m}_{\mathrm{IR}} + \f{R}{2\pi\alpha'}\int_{r_{\mathrm{IR}}}^{r_0}\!r^{\alpha-\beta-1}\,dr\,.
\end{align}
In order that this $\ti{m}_Q$ becomes large for large $r_0$,
we need
\begin{align}
\label{eq:constraint:alpha-beta}
  \alpha - \beta \ge 0\,.
\end{align}

\subsubsection{Current quark mass}
\label{ch:constraints:currentmass}
We need that 
the current quark mass $m_Q$ is large but non-divergent.
According to the dictionary of holography, 
it is given by the mass of a string connecting the D7-brane and 
$\theta\equiv \pi/2$ hypersurface on the boundary. 
That is,
\begin{align}
\label{eq:constraints:currentmass}
  m_{Q} 
  = \lim_{r\to\infty}\f{1}{2\pi\alpha'}\s{-g_{tt}g_{\theta\theta}}\,\l(\f{\pi}{2}-\Theta(r)\r)
  = \f{R}{2\pi\alpha'}\lim_{r\to\infty}r^{\alpha-\beta+\delta}\l(\f{\pi}{2}-\Theta(r)\r).
\end{align}
In order that this is a finite quantity, we need
\begin{align}
\label{eq:constraints:orderofTheta}
  \f{\pi}{2}-\Theta(r)\sim r^{-\alpha+\beta-\delta}\,,
\end{align}
at $r\to\infty$.

\subsubsection{Heavy quark condition}
\label{ch:constraints:heavyquark}
Let us assume a very large $\ti{m}_Q$ (i.e., a large $r_0$),
and introduce a new radial coordinate
\begin{align}\label{eq:s}
  s = \f{r}{r_0}.
\end{align}
By using this, the action $S_\Theta$ \eqref{eq:STheta} for $\Theta(r)$
is rewritten as
\begin{align}\label{eq:Theta:s}
  S_{\Theta} \propto 
  -\int\!ds\,s^{4\alpha-4\beta+4\gamma+3\delta-1}\sin^3\Theta
  \s{1+r_0^{2\delta}(\Theta')^2}\,,
\end{align}
where a prime ($'$) stands for a derivative by $s$.
If $\delta < 0$ or $>0$, the square root becomes
$1$ or $\propto \Theta'$ respectively for large $r_0$ limit.
In both cases, the equation of motion says that $\Theta$ is a constant,
for which \eqref{eq:constraints:orderofTheta} implies
  $\delta = -\alpha + \beta$
which in turn leads to $\delta<0$ from \eqref{eq:constraint:alpha-beta}.
In this case, however, the solution is $\Theta\equiv\pi/2$
and then the current quark mass $m_Q$ becomes $0$ at $r_0\to\infty$ limit.
Therefore the cases of $\delta\ne 0$ are totally excluded.

Now we take $\delta=0$. 
Then \eqref{eq:Theta:s} becomes
\begin{align}
  S_{\Theta} \propto 
  -\int\!ds\,s^{4\alpha-4\beta+4\gamma-1}\sin^3\Theta
  \s{1+(\Theta')^2}\,,
\end{align}
whose equation of motion is equivalent to \eqref{eq:eom:L}.
From the transformation \eqref{eq:rholambda:transf}, 
the current quark mass condition \eqref{eq:constraints:orderofTheta}
implies
\begin{align}
  L \sim r^{-\alpha+\beta+1}\sim \rho^{-\alpha+\beta+1}\,.
\end{align}
Therefore, comparing with \eqref{eq:L:largerho}, we need
\begin{align}
  \kappa_+ = -\alpha+\beta+1\,,
\end{align}
and by using \eqref{eq:kappa} we obtain
\begin{align}
\label{eq:gamma_c}
  \gamma = -\f{3}{4}\l(\alpha-\beta - \f{1}{\alpha-\beta}\r)\,.
\end{align}

\subsection{Bulk Spacetime Curvature}
\label{app:Curvature}
From the string-frame metric \eqref{eq:metric:str} with $\delta=0$ \eqref{eq:const:delta},
we can compute some scalar quantities composed by the curvature, as
\begin{align}
  \cR &= -\frac{4}{R^2} \left(5 \alpha ^2-8 \alpha  \beta +5 \beta ^2-5\right) r^{2 \beta }\,,\\
  \cR_{MN}\cR^{MN} &= \frac{16}{R^4} \left(4 \alpha ^2 (\alpha -\beta )^2+\alpha ^2 (\alpha +\beta )^2+5 \left(\alpha  \beta -\beta ^2+1\right)^2\right) r^{4 \beta }\,,\\
  \cR_{MNPQ}\cR^{MNPQ} &= \frac{8}{R^4} \left(5 \alpha ^4+4 \alpha ^3 \beta +12 \alpha ^2 \beta ^2+5 \left(\beta ^2-1\right)^2\right) r^{4 \beta }\,.
\end{align}
They lead to the condition
\begin{align}
\label{eq:strcurv_cond}
  \beta \le 0\,,
\end{align}
to avoid divergences of these quantities at $r\to\infty$.
The Einstein-frame curvatures are obtained 
simply by replacing $(\alpha,\beta)$ by $(\ti\alpha,\ti\beta)$ in the above
(apart from overall coefficients), 
then the corresponding no-divergence condition is
\begin{align}
\label{eq:Eincurv_cond}
  \ti\beta \le 0\,, \quad\text{i.e.,}\quad \beta\le \gamma\,.
\end{align}
Under the UV weakly-coupled condition \eqref{eq:positivegamma},
the condition \eqref{eq:strcurv_cond} is always stronger than \eqref{eq:Eincurv_cond}.

\section{Equations of Motion for General $\bm{\delta}$}
\label{sec:EoM:general}

In this appendix, we deal with the equations of motion for 
the DBI action \eqref{eq:DBI}
with general value of $\delta$.

First we consider the static configuration \eqref{eq:setup:D7background}.
Substituting into \eqref{eq:background} and \eqref{eq:DBI},
the action becomes
\begin{align}
\label{eq:STheta}
  S_{\Theta} 
  = -\f{2\pi^2T_7R^4V_4}{g_0}\int\!\!dr\,
  r^{4\alpha-4\beta+4\gamma+3\delta-1}
  \sin^3\Theta\s{1+r^{2(\delta+1)}\dot{\Theta}^2}\,,
\end{align}
where
a dot (\,$\dot{}$\,) represents a derivative by $r$ here.
This leads to the equation of motion for $\Theta(r)$,
\begin{align}
  r^{2\delta}\partial_r\l(\f{\dot{\Theta}\sin\Theta}{\s{1+r^{2(\delta+1)}\dot{\Theta}^2}}\r)
  - (4\alpha-4\beta+4\gamma+5\delta)r^{2\delta-1}
  \f{\dot{\Theta}\sin\Theta}{\s{1+r^{2(\delta+1)}\dot{\Theta}^2}}
&\nonumber\\
  + 3\cos\Theta\sin^2\Theta\s{1+r^{2(\delta+1)}\dot{\Theta}^2}
&\quad
  = 0\,.
\end{align}
Since we want to consider turning-around configurations,
we set the IR boundary condition for $\Theta(r)$ as
\begin{align}
  \Theta(r_0) = 0\,,
\qquad
  \dot{\Theta}(r_0) = \infty\,.
\end{align}

On this configuration, we consider small linear fluctuations of $\psi$ and $A_a$ in the form of
\begin{align}
  \delta A_{x} &\sim e^{-ik_\mu x^\mu}Y_{\ell mn}(\Omega_3)a(r)\,,
\\
  \delta \psi &\sim e^{-ik_\mu x^\mu}Y_{\ell mn}(\Omega_3)\Psi(r)\,,
\end{align}%
where the regularity at $r=r_0$ requires the IR boundary conditions 
\begin{align}
  \dot{a}(r_0) = 0\,,
\qquad
  \dot{\Psi}(r_0) = 0\,.
\end{align}
For these, the leading (quadratic) fluctuation terms of the DBI action read,
respectively,
\begin{align}
\label{eq:EoM:Sa}
  S_{a} 
  &= -\f{4\pi^4\alpha'^2T_7R^2V_4}{g_0}\int\!dr\,r^{-4\beta+4\gamma+\delta-1}
  \f{\sin\Theta}{\s{1+r^{2(\delta+1)}\dot{\Theta}^2}}
\nonumber\\
  &\qquad\quad\times
  \l[\l\{\ell(\ell+2)r^{2(\alpha+\beta)}-M^2R^2r^{2(\delta+1)}\sin^2\Theta\r\}\l(1+r^{2(\delta+1)}\dot{\Theta}^2\r)a^2
  + r^{2(\alpha+\beta+\delta+1)}\sin^2\Theta\,\dot{a}^2
\r]\,,
\\
\pagebreak[3]
\label{eq:EoM:SPsi}
  S_{\Psi} 
  &= -\f{\pi^2T_7R^4V_4}{g_0}\int\!dr\,r^{2\alpha-6\beta+4\gamma+3\delta-1}
  \f{\cos^2\Theta\sin\Theta}{\s{1+r^{2(\delta+1)}\dot{\Theta}^2}}
\nonumber\\
  &\qquad\quad\times
  \l[\l\{\ell(\ell+2)r^{2(\alpha+\beta)}-M^2R^2r^{2(\delta+1)}\sin^2\Theta\r\}\l(1+r^{2(\delta+1)}\dot{\Theta}^2\r)\Psi^2
  + r^{2(\alpha+\beta+\delta+1)}\sin^2\Theta\,\dot{\Psi}^2
\r]\,.
\end{align}
We notice that \eqref{eq:EoM:SPsi} and \eqref{eq:EoM:Sa} are very similar
to each other,
and the only difference is the factor of 
$r^{2(\alpha-\beta+\delta)}\cos^2\Theta$ in the integrand.

\section{Schr\"odinger Form of the Equations of Motion}
\label{sec:Schroedinger}
The linear equations \eqref{eq:eom:A} and \eqref{eq:eom:psi} can be
rewritten as 
the conventional forms of Schr\"odinger equations,
\begin{align}
\label{eq:Schroedinger:A}
  \l(-\partial_\chi^2+V_a(\chi)\r)\ti{a} &= (M_aR)^2\,\ti{a}\,,\\
\label{eq:Schroedinger:psi}
  \l(-\partial_\chi^2+V_\Psi(\chi)\r)\ti{\Psi} &= (M_\Psi R)^2\,\ti{\Psi}\,,
\end{align}
by proper transformations of functions and coordinates.
For $\ell=0$,
the transformations are given by
\begin{align}
  &\partial_\chi\rho = \f{1}{\s{\ti{P}_0}}\,,
  \qquad
  \ti{P}_0 \equiv \f{P_0}{M^2R^2} 
  = \f{1+\dot{L}^2}{(\rho^2+L^2)^{\alpha+\beta+1}}\,,\\
  &\ti{a} = f_aa\,,
  \quad
  \rho'' + 2(\rho')^2\partial_\rho\log{f_a} = \f{P_1}{\ti{P}_0}\,,\\
  &\ti{\Psi} = f_\Psi\Psi\,,
  \quad
  \rho'' + 2(\rho')^2\partial_\rho\log{f_\Psi} = \f{P_1+Q_1}{\ti{P}_0}\,,
\end{align}
which lead to the potentials
\begin{align}
  V_i = \f{\rho''\dot{f}_i + 2(\rho'')^2\ddot{f}_i}{f_i}
  \qquad (i=a,\Psi)\,,
\end{align}
where a prime (${}'$) stands for a derivative by $\chi$.
By using these equations, 
we can draw the shapes of the potentials $V_a(\chi)$ and $V_\Psi(\chi)$ numerically.

Furthermore, under the condition \eqref{eq:gamma_c}, 
$Q_1(\rho)/P_1(\rho)$ decays at $\rho\to\infty$ and so
the equations for $a$ and $\Psi$ have the same asymptotics.
By using \eqref{eq:largerho:a0a1b1}, they can be solved as
\begin{align}
  \chi &\simeq \chi_\infty - \f{\rho^{-(\alpha+\beta)}}{\alpha+\beta}\,,
\qquad
  f \simeq f_0\rho^{\f{\zeta-(\alpha+\beta)}{2}}\,,\\
  V &\simeq \f{\zeta^2-(\alpha+\beta)^2}{4}\rho^{2(\alpha+\beta)} \nonumber\\
  &\simeq \f{\zeta^2-(\alpha+\beta)^2}{4(\alpha+\beta)^2}
  \f{1}{(\chi_\infty-\chi)^2}
\,.
\end{align}
Note that the null energy condition \eqref{eq:NEC:2} guarantees that 
the coefficient is positive.
Therefore, 
there is an infinitely high potential barrier at $\chi=\chi_\infty$,
which ensures a stable, discrete spectrum.
At that time, the leading behavior of the normalizable/non-normalizable solutions for the Schr\"odinger equation \eqref{eq:Schroedinger:A} are
\begin{align}
  \ti{a}\,,\ti{\Psi} \sim \l(\f{1}{\chi_\infty-\chi}\r)^{\f{1}{2}\l(-1\pm\f{\zeta}{\alpha+\beta}\r)}\,,
\end{align}
or equivalently,
\begin{align}
  a\,,\Psi \sim \rho^{-\zeta}
\;\;\text{(normalizable)}\,,
  \qquad
  a\,,\Psi \sim 1
\;\;\text{(non-normalizable)}\,.
\end{align}
The former is our desirable solution,
and we need to choose proper energy eigenvalues
to make the coefficient of the latter solution to be zero.
This is 
equivalent to the procedure
we carried out in \S\ref{ch:Numerical_results}.

\bibliographystyle{JHEP}
\bibliography{reference}

\end{document}